\documentclass[useAMS,usenatbib]{mnras}
\usepackage{graphicx}
\usepackage{natbib}
\usepackage{amssymb}
\usepackage{setspace}
\usepackage{amssymb}
\usepackage{epsfig,color}
\usepackage{multicol}
\usepackage{longtable}
              
\long\def\symbolfootnote[#1]#2{\begingroup%
\def\thefootnote{\fnsymbol{footnote}}\footnote[#1]{#2}\endgroup} 

\newcommand{\Msol}{~$M_{\odot}$} 
\newcommand{\sqc}{cm$^{-2}$}                   
\newcommand{\cc}{cm$^{-3}$}                    

\newcommand{\Gauss}{{\sc Gaussclumps}}  
\newcommand{\Dendrogram}{{\sc Dendrogram}}  
\newcommand{\Herschel}{{\it Herschel}}  
                                                                                                                  
\newcommand{\twelveCO}{$^{12}$CO}  
\newcommand{\thirthCO}{$^{13}$CO}  
\newcommand{\CeightO}{C$^{18}$O}
\newcommand{\Hii}{H\,{\sc ii}~}

\title[Clump populations in Rosette]{Spatially associated clump populations in Rosette from CO and dust maps}

\author[Veltchev et al.]
{	
\parbox{\textwidth}{Todor V.~Veltchev$^{1,2\,\star}$, Volker Ossenkopf-Okada$\,^3$, Orlin Stanchev$^1$, Nicola Schneider$^3$, Sava Donkov$^4$, and  Ralf S. Klessen$^2$}\vspace{0.4cm} \\
\parbox{\textwidth}{
  $^1$University of Sofia, Faculty of Physics, 5 James Bourchier Blvd., 1164 Sofia, Bulgaria\\
  $^2$Universit\"at Heidelberg, Zentrum f\"ur Astronomie, Institut f\"ur Theoretische Astrophysik, Albert-Ueberle-Str. 2, 69120 Heidelberg, Germany\\
  $^3$I. Physik. Institut, University of Cologne, D-50937 Cologne, Germany\\
  $^4$Department of Applied Physics, Technical University, 8 Kliment Ohridski Blvd., 1000 Sofia, Bulgaria }
}

\date{Submitted 2017 Xxxxx XX}

\pagerange{\pageref{firstpage}--\pageref{lastpage}} \pubyear{2017}

\begin{document}
\label{firstpage}
\maketitle

\begin{abstract}
Spatial association of clumps from different tracers turns out to be a valuable tool to determine the physical properties of molecular clouds. It provides a reliable estimate for the $X$-factors, serves to trace the density of clumps seen in column densities only and allows to measure the velocity dispersion of clumps identified in dust emission. We study the spatial association between clump populations, extracted by use of the \Gauss{} technique from \twelveCO\,($1-0$), \thirthCO\,($1-0$) line maps and \Herschel~dust-emission maps of the star-forming region Rosette, and analyse their physical properties. All CO clumps that overlap with another CO or dust counterpart are found to be gravitationally bound and located in the massive star-forming filaments of the molecular cloud. They obey a single mass-size relation $M_{\rm cl}\propto R_{\rm cl}^\gamma$ with $\gamma\simeq3$ (implying constant mean density) and display virtually no velocity-size relation. We interpret their population as low-density structures formed through compression by converging flows and still not evolved under the influence of self-gravity. The high-mass parts of their clump mass functions are fitted by a power law ${\rm d}N_{\rm cl}/{\rm d}\,\log M_{\rm cl}\propto M_{\rm cl}^{\Gamma}$ and display a nearly Salpeter slope $\Gamma\sim-1.3$. On the other hand, clumps extracted from the dust-emission map exhibit a shallower mass-size relation with $\gamma=2.5$ and mass functions with very steep slopes $\Gamma\sim-2.3$ even if associated with CO clumps. They trace density peaks of the associated CO clumps at scales of a few tenths of pc where no single density scaling law should be expected.  
\end{abstract}


\section{Introduction}
Observations of star-forming (SF) regions reveal the complex structure of the associated molecular clouds (MCs). The latter have been originally resolved into clumps of sub-parsec size. Recent high-resolution observations (e.g. with {\it Herschel}) show that the dense gas in MCs is predominantly concentrated in filamentary networks which probably play a central role in the star-formation process through further subfragmentation into dense prestellar cores with sizes below $0.1$~pc and densities up to $10^6$~\cc \citep{Andre_ea_14}. 
A number of clump extraction techniques have been developed in the last three decades. They have been applied initially to molecular-line emission and dust-extinction maps and, later, to dust-emission maps which enabled in-depth studies of numerous SF regions in the Solar neighbourhood. As a major result, the obtained velocity-size and mass-size relationship for MC fragments and clumps emphasise the role of turbulence and gravity in the cloud evolution and their interplay \citep{Larson_81, Solomon_ea_87, Heyer_Brunt_04, Heyer_ea_09}. Virial analysis of clump populations hinted at possible mechanisms of clump formation \citep[e.g.][]{Myers_Goodman_88, BM_92, Dib_ea_07}. It also turned out that mass functions of dense clumps resemble the stellar initial mass function (\citealt{ALL_07, Andre_ea_10}; see, however, \citealt{Clark_ea_07}). An open issue remains whether the extracted clumps represent real, distinct physical entities.

Clump-finding algorithms have been put to test with the advance of numerical simulations of MC evolution. \citet{BP_MacLow_02} and \citet{Shetty_ea_10} investigated the role of projection effects on the clump properties and their virial analysis, while \citet{Beaumont_ea_13} estimated the uncertainties of the derived characteristics of \thirthCO~clumps. An important clue to link clump properties with the physics of star formation is provided through the derivation of a clump mass function \citep{SG_90, WBS_95, Kramer_ea_98, Heithausen_ea_98} and the interpretation of its parameters like slope and characteristic mass \citep[e.g.][]{ALL_07, VDK_13}. Clump properties depend essentially on how the entire population is considered: as a set of independent entities or as a hierarchy in the position-position-position (PPP) / position-position-velocity (PPV) space. A widely used technique for non-hierarchical clump extraction is {\sc Clumpfind} \citep{Williams_ea_94}. It is based on eye inspection and identifies each peak on the intensity map with a clump. Recently \citet{Menshchikov_ea_12} proposed another non-hierarchical algorithm {\sc getsources} which aims at 2D image decomposition in continuum maps at multiple scales and wavelengths and is appropriate for detection of clumps in crowded regions.

On the other hand, defining clumps as hierarchical objects reflects the fractal structure of MCs \citep{Elmegreen_Falgarone_96, Elmegreen_02}. This allows to link their properties to the general physics of SF regions as described by the above-mentioned scaling relations of mass and velocity. One hierarchical method for clump delineation is the \Dendrogram~technique \citep{Rosolowsky_ea_08} which traces the segmentation of cloud structures as one increases the threshold intensity. 

Yet projection effects can be misleading in studies of cloud hierarchy. One can reduce them by the use of the clump-extraction technique \Gauss~\citep{SG_90, Kramer_ea_98}. It allows for a distinction of multiple coherent structures along the same line of sight by decomposing any structure into PPV or PPP Gaussian clumps which may overlap on the sky maps. For all smaller clumps in the resulting clump hierarchy the method derives significantly different properties from those that {\sc Clumpfind} does \citep{Schneider_Brooks_04}. By allowing for the superposition of nested Gaussian clumps to form a large structure \Gauss~also enables the retrieval of the hierarchical structure of a MC. Therefore, the approach of \Gauss~is neither purely hierarchical, like the \Dendrogram~technique, nor purely non-hierarchical, like {\sc Clumpfind}, but inherits advantages of both approaches. It characterises confined entities, but it also allows for the detection and analysis of nested structures from a single or multiple tracer(s).  
Our main goal in this work is to show what could be learned from spatial association of rich clump samples obtained via \Gauss~on molecular-line (\twelveCO,~\thirthCO) and dust-emission maps. 

The Rosette star-forming region is appropriate for this purpose -- it has been intensively investigated in the last decades and mapped at different wavelengths with high angular resolution. Clumpy structures in the MC complex have been studied using various algorithms and tracers \citep{WBS_95, Schneider_ea_98, Dent_ea_09, DiFran_ea_10}. The statistical analysis of the highly filamentary structure of the complex and of the location of young stellar objects provides insight to the nature of the star-formation efficiency in individual clumps and it showed that the star formation in the cloud is not driven by radiative feedback \citep{Schneider_ea_12}. 

The paper is organised as follows. First, the used observational data are presented (Section \ref{Observational data}). Section \ref{Clump extraction} reviews the algorithm and the calculation of clump sizes and masses. The adopted criterion for spatial association of clump populations and their statistics in Rosette is described in Section \ref{Clump association}. Section \ref{Physical analysis} contains results from the performed physical analysis of the clump samples: size distributions and the relationships size vs. mass and mass vs. virial parameter. The clump mass functions are derived and studied in Section \ref{CMFs}. Some problems and uncertainties are discussed in Section \ref{Discussion}. A summary of this work is given in Section \ref{Summary}.

\section{Observational data on Rosette MC}
\label{Observational data}

\subsection{Selected region}
\begin{figure*} 
\begin{center}
\includegraphics[width=1.\textwidth, keepaspectratio]{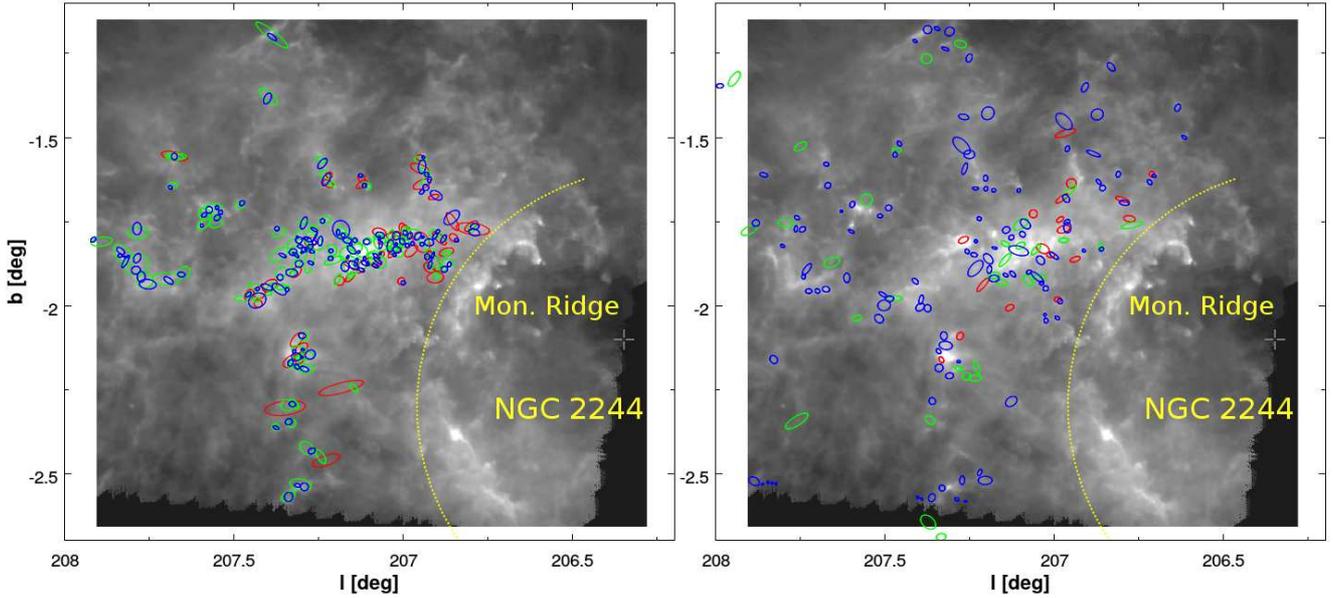}
\vspace{0.3cm}  
\caption{Locations of the extracted associated (left) and non-associated (right) clumps from \twelveCO~(red), \thirthCO~(green) and dust emission (blue), imposed on a {\it Herschel} map of Rosette (combined SPIRE 250 + 350 + 500 $\mu$m; available from the ESA Sky Archive http://sky.esa.int). The boundary of the region excluded from consideration is marked with dashed line (see text).}
\label{fig_assoc_map}
\end{center}
\end{figure*}

In this study, we consider the Rosette molecular cloud (hereafter, RMC) with filamentary structures connected to it and traced on {\it Herschel} column-density maps \citep{Motte_ea_10, Hennemann_ea_10, DiFran_ea_10, Schneider_ea_10, Schneider_ea_12}, excluding the zone of direct interaction between the molecular cloud and the expanding \Hii region around NGC\,2244, the so called `Monoceros Ridge' \citep{Blitz_Thaddeus_80} (Fig. \ref{fig_assoc_map}). The physical properties of the clump/core population in the Monoceros Ridge can be significantly influenced by the effects of stellar feedback such as gas compression by the expanding ionisation front or heating by the radiation from the OB cluster NGC\,2244. \citet{Motte_ea_10} detected more massive dense cores forming in this zone while \citet{Schneider_ea_12} and \citet{Cambresy_ea_13} showed that there is no indication for large-scale triggering of star-formation further inside Rosette cloud. 

Thus the region studied in this paper is restricted to zones where the star formation activity is probably not caused by direct external gas compression. A distance of $1330$~pc to the region was adopted \citep{LAL_11} though distances up to $1.6$~kpc were also used in the literature \citep[see][]{RZ_Lada_08}. The choice of distance value does not affect the analysis of scaling relations, the virial analysis and the derivation of the slopes of clump mass functions provided in the next sections.

\subsection{Molecular-line and dust emission maps}
\label{Maps}

We use maps of the Rosette star-forming region in \twelveCO\,$(J=1-0)$ (hereafter, \twelveCO) and \thirthCO\,$(J=1-0)$ (hereafter, \thirthCO) emission taken with the 14m telescope of Five College Radio Astronomy Observatory (FCRAO). These data sets were presented and discussed by \citet{Heyer_ea_06}.
The beam FWHM of the CO data is $\sim$46\arcsec, the spectral resolutions are 0.127 km s$^{-1}$ (\twelveCO) and 0.133 km s$^{-1}$ (\thirthCO). 
All temperatures are given on the main beam brightness temperature scale. 

The original maps of dust emission were obtained from \Herschel~observations at four wavelengths of PACS and SPIRE:
160, 250, 350 and 500 $\mu$m \citep[see][for details]{Schneider_ea_10, Schneider_ea_12}. The maps were optimised for {\it extended} emission\footnote{In contrast to optimized maps for point sources.} through the standard reduction methods in the HIPE pipeline and its scripts. For example, gains for extended emission are applied as {\tt applyExtendedEmissionGain} was set to {\tt TRUE} in the HIPE SPIRE pipeline. In addition, the {\it Planck} offsets for a zero-point correction were applied. The maps were convolved to a common angular resolution of 36\arcsec. 

Contribution of the fore- or background in Rosette is not as significant as it is in other star-forming regions located close to or deep in the Galactic plane. Rosette is rather isolated which is also well reflected in velocity -- only velocity structures in this region are detectable along the line of sight. In our view, possible overestimations of column density and, hence, of mass due to presence of fore- or background structures could hardly reach more than a few percent (in individual small subregions). Therefore we abstained from applying any fore- or background corrections to the {\it Herschel} maps. The major uncertainty of masses should be due to projection effects within the RMC, as specified in Sect. \ref{Estimation of X-factor}.

In a next step, a pixel-to-pixel grey-body fit to the data was performed, assuming that dust opacity per gas mass follows a power-law in the far infrared:
\begin{equation}
\kappa_\nu \, = \,0.1 \,\Big(\frac{\nu}{1000~{\rm GHz}}\Big)^b~~\frac{{\rm cm}^{2}}{\rm g}~,
\end{equation}
with $b\equiv2$ and leaving dust temperature and column density as free parameters. We assume here for dust emissivity the value for dense clouds used, e.g., by \citet{Hill_ea_11}, \citet{Russeil_ea_13}, \citet{Roy_ea_13}. It corresponds to a dust opacity of $2.3\times 10^{-25} \mathrm{cm}^2/$H at $\nu=1$~THz ($\lambda=300$~$\mu$m), when a mean mass $\mu_{\rm H,\,Gal}= 1.37m_{\rm u}$ per hydrogen (with $m_{\rm u}$ as the atomic mass unit) is adopted. The various dust models from \citet{Ossenkopf_Henning_94} cover opacities between $1.0\times 10^{-25} \mathrm{cm}^2/$H and $3.0\times 10^{-25} \mathrm{cm}^2/$H when assuming a gas-to-dust ratio of 100. The lower limit rather represents non-coagulated dust in diffuse clouds so that it is probably not representative for Rosette. For {\em Planck} cold cores \citet{Juvela2015} estimated $1.5\times 10^{-25} \mathrm{cm}^2/$H when converted to 300~$\mu$m. Taking this range, we estimate that the dust emissivity has an accuracy of $30-50$\%. We also tested fits using lower exponents for the dust spectral index $1.5\le \beta<2$, but found that the uncertainty in the resulting column density is smaller than the one governed by the absolute emissivity. Thus we estimate that the dust-based column-density map of Rosette has an accuracy of about $50$\%, limited mainly by the uncertainty of the dust emissivity in the cloud and, to a lower degree, by the uncertainty in the SED-fitting procedure (constant line-of-sight temperature approximation). 

\section{Clump extraction and characteristics}
\label{Clump extraction}

In this Section we comment on the adopted settings of the chosen clump-extraction technique and on the calculation of sizes and masses of clumps from molecular-line and dust emission maps.

\subsection{Overview of the algorithm}
\label{Gaussclumps technique}
The algorithm \Gauss~was originally developed for the iterative decomposition of three-dimensional intensity distributions into individual clumps assuming a Gaussian shape \citep{SG_90}. However, it can be also applied to dust continuum maps \citep[e.g.][]{Motte_ea_03}. \Gauss~is implemented in two widely used software packages: {\sc Gildas}\footnote{https://www.iram.fr/IRAMFR/GILDAS/} and {\sc Cupid}\footnote{http://starlink.eao.hawaii.edu/starlink/CUPID}. We used the {\sc Gildas} implementation for clump extraction on the \twelveCO~and \thirthCO~emission maps and the {\sc Cupid} implementation for the {\it Herschel} data. 

In both cases the choice of \Gauss~control parameters is important. In general, when the three ``stiffness'' parameters are large ($s_a = s_b = s_c =50$), part of the noise peaks will be interpreted as clumps while adoption of very low values ($s_a = s_b = s_c = 0.01$) excludes a significant fraction of the emission above the Gaussian noise from the clump composition procedure. We set $s_a = s_b = s_c = 1.0$ following \citet{Kramer_ea_98}\footnote{Somewhat larger values of $s_b$ are also applicable; cf. \citealt{Schneider_ea_98}.}. The threshold below which the algorithm stops iterating is usually taken to be several times the data rms \citep{Curtis_Richer_10, Pekruhl_ea_13, Csengeri_ea_14}. We adopt a stringent threshold of 5 times the noise level for the input \twelveCO~and \thirthCO~maps. Dealing with the {\it Herschel} map, several test runs with different thresholds were made. A standard significance threshold of $3$ times the noise level yielded clumps with sizes $0.08~{\rm pc}\le R_{\rm cl} \le 0.47$~pc and masses $M_{\rm cl}$ between 2 and 268 \Msol, whereas the mass range $M_{\rm cl}\lesssim 10$~\Msol~was severely underpopulated\footnote{For definitions of clump size and mass, see next Section.}. To increase the statistics of dense dust cores and the number of corresponding associations, we lowered the threshold stepwise. The best test value turned out to be $1.5\sigma$: the minimal $R_{\rm cl}$ shifted downwards just by $0.02$~pc, while many objects with characteristics of dense cores ($0.1\lesssim R_{\rm cl}/1~{\rm pc}\lesssim0.3$, $M_{\rm cl}\le10$~\Msol) appeared. Thus the total number of dust clumps increased by a factor of 2.5, while the number of associations (as defined in Sect. \ref{Method for clump association}) increases by a factor of two. As the \twelveCO~and \thirthCO{} clumps occupy only a small fraction of the map (see Fig.~\ref{fig_assoc_map}) this close correspondence indicates that most associated dust clumps are real. The adopted $1.5\sigma$ threshold limit to detect their peaks corresponds to a column density $N_{\rm lim}(\rm dust)=1.9\times10^{21}$~\sqc.

\subsection{Sizes and masses of Gaussian clumps}
\label{Clump parameters}

The effective diameter of a Gaussian clump, extracted from a molecular-line emission map, is calculated as the geometrical mean of its major and minor axes on the sky plane (position-position space):
\[ D_{\rm cl}=\sqrt{\Delta x_{\rm cl}\,\Delta y}_{\rm cl}~~, \] 
where $\Delta x_{\rm cl}$ and $\Delta y_{\rm cl}$ are defined as beam-deconvolved full widths of half-maxima (FWHM) of the fitted Gaussian curves along those directions, converted to linear sizes. Hereafter, we label `clump size' the half of the effective diameter $R_{\rm cl}=0.5 D_{\rm cl}$. Our observational data allow for linear resolutions are of $0.15$~pc (CO maps) and $0.12$~pc (dust maps), respectively. 

The masses of the CO clumps were derived through three-dimensional integration of the brightness temperature $T_{\rm b}$ over the PPV space:
\[  M_{\rm cl}{\rm (CO)} = \mu_{\rm mol}X\int\!\!\!\!\int\!\!\!\!\int T_{\rm b}\,dxdydv~~~~~~~~~~~~~~~~~~~~~~~~~~~~~~~~~~~~~~~~~~~ \]
\[  = \mu_{\rm mol}XT_{\rm b,\,0} \int\!\!\!\!\int\!\!\!\!\int e^{-\frac{(x-x_0)^2}{2\sigma_x^2}}\!e^{-\frac{(y-y_0)^2}{2\sigma_y^2}}\!e^{-\frac{(v-v_0)^2}{2\sigma_v^2}}\,dxdydv  \]
\begin{equation}
 \label{eq_mass_CO_clumps}
 = \Big( \frac{1}{2}\sqrt{\frac{\pi}{\ln 2}}\Big)^3 \mu_{\rm mol} X T_{\rm b,\,0}\,\Delta x_{\rm cl}\,\Delta y_{\rm cl}\,\Delta v_{\rm cl}~~~~~~~~~~~~~
\end{equation}
where $T_{\rm b,\,0}$ is the peak value of $T_{\rm b}$ within the given clump, $(x_0, y_0, v_0)$ are the positions of its centre in the PPV space, $\Delta v_{\rm cl}=2\sigma_{\rm cl}\sqrt{2\ln2}$ is the FWHM in velocity and $\mu_{\rm mol}=2.74 m_{\rm u}$ is the mean particle mass in molecular gas. The adopted estimates of the conversion factors $X$ from integrated CO (\twelveCO~or~\thirthCO) intensity to hydrogen column density $N$ are discussed in Sect. \ref{Estimation of X-factor}. 

Formula (\ref{eq_mass_CO_clumps}) was used by \citet{SG_90} to derive masses of Gaussian clumps, extracted from maps of an optically thin tracer (\CeightO). Why it is applicable also to \twelveCO~and \thirthCO{} clumps in the Rosette region? Optical depth of \thirthCO{} $(J=1-0)$ in various zones in RMC was assessed by \citep{Schneider_ea_98} as several intensity ratios have been modelled with external UV radiation. This line might be optically thick in the Monoceros ridge, at the position of the infrared source AFGL 961 and in the center zone $(l\sim207.\!^{\circ}1,~b\sim-1.\!^{\circ}8)$, i.e. for a few positions, in the vicinity of very high column density peaks. The Monoceros ridge (cf. Fig.~\ref{fig_assoc_map}) and the clumps identified with AFGL 961 were excluded from our consideration whereas a few ones located in the centre zone would hardly affect the results presented in the next sections. For the remaining zones of lower (column) density gas, it can be assumed that \thirthCO{} line is predominantly optically thin. The latter was shown also by \citet{WBS_95} whose selected region in Rosette largely coincides with ours (see their Fig. 1).

In regard to the Gaussian clumps, extracted from maps of the optically thick \twelveCO{} $(1-0)$ emission, \citet{Kramer_ea_98} found that their mass is still proportional to $X T_{\rm b,\,0}\,\Delta x_{\rm cl}\,\Delta y_{\rm cl}\,\Delta v_{\rm cl}$, ``although the applicability of the $X$-factors from the literature to individual clumps may be questionable'' (Sect. 3.2.2 therein). In this study we work with an {\it averaged} $X$-factor, derived from associations of \twelveCO/\thirthCO{} clumps with dust counterparts (i.e. extracted from an optically thin emission, Sect. \ref{Estimation of X-factor}). Therefore the \twelveCO{} clump mass, calculated from equation (\ref{eq_mass_CO_clumps}) could be underestimated -- in the worst case -- by some constant factor. However, this would not affect neither the slopes of the studied scaling relations, nor the virial analysis since all \twelveCO{} clumps have been found to be gravitationally bound (Sect. \ref{Virial_analysis_M}).

The masses of the dust clumps, extracted from the {\it Herschel} map, were obtained in analogous way to equation (\ref{eq_mass_CO_clumps}), through two-dimensional integration of column density:

\begin{equation}
 \label{eq_mass_dust_clumps}
 M_{\rm cl}{\rm (dust)}=\Big( \frac{1}{2}\sqrt{\frac{\pi}{\ln 2}}\Big)^2 \mu_{\rm H,\,Gal} N_0\,\Delta x_{\rm cl}\,\Delta y_{\rm cl}~~,
\end{equation}
where $N_0$ is the peak value of column density within the clump and the adopted mean particle mass $\mu_{\rm H,\,Gal}= 1.37m_{\rm u}$ is representative for Galactic abundances of atomic and molecular hydrogen and other elements \citep{Draine_11}.

\begin{figure*} 
\includegraphics[width=1.\textwidth]{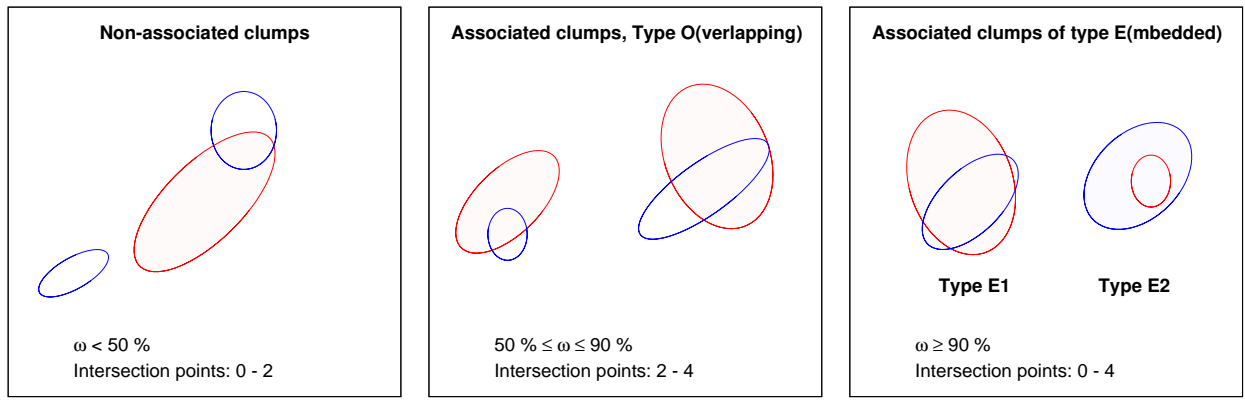}
\vspace{0.25cm}  
\caption{Criteria for association of Gaussian clumps from Population 1 (red) and Population 2 (blue), based on the value of the overlap coefficient $\omega$ (see text): non-associated (left), overlapping (middle) and embedded (right) clumps.}
\label{fig_criterion_clump_association}
\end{figure*}



\section{Spatial association between clump populations}
\label{Clump association}
Cross-identification (hereafter, association) of clumps from different tracers could be used as a tool to determine the physics of structures in MCs. In this Section we present a method to associate Gaussian clumps, introduce two types of associated clumps and provide statistics of the associated populations in the RMC.

\subsection{Association between Gaussian clumps}
\label{Method for clump association}
Various tracers of cloud structure are sensitive to different ranges of density and optical depth. While the $J=1\rightarrow0$ lines of \twelveCO~and \thirthCO~are typical tracers of molecular gas with densities $n\gtrsim10^2$~\cc~and $n\gtrsim 10^3$~\cc, respectively, the far-infrared and submillimetre emission of dust enables one to trace the cloud structure from very low densities up to its dense cores ($n>10^5$~\cc). Therefore it is instructive to perform association between the extracted clump populations and to compare their locations and physical characteristics. 

Since the projections of CO PPV clumps onto the celestial plane as well as the 2-dimensional clumps extracted from \Herschel~maps through \Gauss~are ellipses, an appropriate criterion to associate clump pairs should be based on the theory of intersecting ellipses. We make use of a method developed by \citet{Hughes_Chraibi_12}. Its output are the number of intersection points and the intersection area of a considered pair of ellipses. In our treatment, the ellipse representing each clump is defined by the half-maximum contour. 

The applied criterion for clump association is illustrated in Fig. \ref{fig_criterion_clump_association}. As one considers a pair of clumps from two samples, Population 1 and Population 2, the {\it overlap coefficient} $\omega$ is defined as the intersection area, normalised to the area of the smaller ellipse. In case of zero or one intersection points the ellipses do not overlap at all ($\omega=0$~\%) or the smaller one is completely embedded into the larger ($\omega=100$~\%). When the intersection points are 2, 3 or 4, the value of $\omega$ is decisive to discriminate between non-associated, overlapping and embedded clumps. To distinguish non-associated from overlapping clumps (Type O) we adopt a conservative range $50$\%$\le\omega<90$~\%.
Type E (embedded) is assigned to clump pairs with $\omega\ge90$~\%; such criterion is strong enough to ensure that most of the mass of a centrally condensed clump is contained in the overlap area. Two subclasses of type E are introduced depending on whether the clump from Population 1 is larger than its associate from Population 2 (Type E1), or vice versa (Type E2). An additional requirement for associating a \twelveCO~and a \thirthCO~clump is that the velocity ranges $(v_0-\Delta v_{\rm cl}/2) \le v\le (v_0+\Delta v_{\rm cl}/2)$ in both tracers contain at least one velocity channel in common. 

\begin{table}
\caption{Statistics of the associated Gaussian clumps in RMC.}
\label{table_clump_association_statistics} 
\begin{center}
\begin{tabular}{ccc}
\hline 
\hline 
Tracer &\multicolumn{2}{c}{Nr. of clumps} \\ 
~ & In total & Associated \vspace{4pt}\\
\hline 
\twelveCO & ~68 & ~{\bf 49} \\
\thirthCO & 130 & ~{\bf 99} \\
Dust & 249 & {\bf 133} \\
\hline 
\hline 
\end{tabular} 
\end{center}
\smallskip 
\end{table}

\begin{table}
\caption{Statistics of the associated pairs in RMC. The last two (one) columns contain the number ratios between the number of associations of different type.}
\label{table_associated_pairs_statistics} 
\begin{center}
\begin{tabular}{c@{~}c@{~~~}c@{~~~}c@{~~}c}
\hline 
\hline 
\multicolumn{3}{c}{Associated pairs}  & \multicolumn{2}{c}{\it Number ratios} \\ 
Pop 1-Pop 2 & Nr.$^{\ast}~$ & Notation in Figs. & E1/O & E2/O \\
~ & ~ & \ref{fig_mass-mass_diagram}, \ref{fig_mass-size_Gaussian_assoc}, \ref{fig_sigma_scaling}, \ref{fig_mass_alpha_vir}, \ref{fig_M-R_vir}, \ref{fig_appendix_R-R_diag} & ~ & ~ \vspace{2pt} \\
\hline 
\twelveCO\,-\,Dust & ~63 & diamonds & 0.80 & 0.03 \\
\thirthCO\,-\,Dust & 112 & squares & 1.07 & 0.07 \\
\twelveCO\,-\,\thirthCO& ~37 & circles & 0.76 & 0.00 \\
\hline 
\multicolumn{5}{c}{{\it Associations with the same population}} \\
 ~ & ~ & ~ & \multicolumn{2}{c}{E/O} \\ 
\hline
\twelveCO\,-\,\twelveCO & 11 & -- & \multicolumn{2}{c}{0.57} \\
\thirthCO\,-\,\thirthCO & 12 & -- & \multicolumn{2}{c}{0.20} \\
Dust\,-\,Dust & ~7 & -- & \multicolumn{2}{c}{1.33} \\
\hline 
\hline 
\end{tabular} 
\end{center}
{\footnotesize$\ast$ Some clumps of Population 1 belong to more than one pair. }
\smallskip 
\end{table}

\subsection{Associated clumps in Rosette}
The associated objects populate the dense, main star-forming region in the RMC and its most pronounced filaments (Fig. \ref{fig_assoc_map}, left; cf. Fig. 1 in \citealt{Schneider_ea_12}). 
In contrast, the non-associated clumps are widely distributed over the cloud (Fig. \ref{fig_assoc_map}, right); the few non-associated \twelveCO~objects occur only at $l<207.\!\!^\circ3$. Most non-associated \thirthCO~clumps are detected in channels of high radial velocity. 

In Tables \ref{table_clump_association_statistics} and \ref{table_associated_pairs_statistics} we provide statistics of the associated clumps from each population. About 75 \% of all CO clumps (\twelveCO~and \thirthCO) have at least one associate in one or both of the other tracers. Considering the dust clump population, the fraction of associated clumps is only 53 \%. Twenty \twelveCO~clumps in total have associates in the other two tracers; several of them are associated with more than one \thirthCO~and/or dust clump. The vast majority of the non-associated dust clumps are small ($R_{\rm cl}<0.3$~pc), low mass ($M_{\rm cl}<10$\Msol) objects; part of them may be artefacts from our low detection limit.  

Among the associated clump pairs (Table \ref{table_associated_pairs_statistics}), Type O and Type E turn out to be equal or close in number. Considering only pairs of Type E, dust clumps are -- in general -- embedded into their CO counterparts and \thirthCO~clumps are embedded into \twelveCO~ones. The exceptions enable us to estimate the $X$-factors (Sect. \ref{Estimation of X-factor}). We also computed the associations for clumps of the same tracer to quantify hierarchies but the statistics of these associated pairs is poor due to the limited dynamic range of clump sizes. Their analysis provides reference values in the study of the mass-size relationship (cf. Sect \ref{Size-mass relationships}). 


\section{Physical analysis}
\label{Physical analysis}

The analysis in this Section is focused on the associated clump populations. Since the calculated clump masses depend crucially on the $X$-factor (cf. Sect. \ref{Clump parameters}), we suggest first an approach to estimate its average value in the RMC (Sect. \ref{Estimation of X-factor}). The derived size-mass relationships are presented in Sect. \ref{Size-mass relationships}. In view of these and of the clump size distributions, we explore the strength of our method to associate individual clumps, investigating cross-correlations between the three maps of Rosette as functions of the spatial scale (Sect. \ref{Cross-correlation between maps}). Then, in Sect. \ref{Virial analysis}, we study the velocity dispersions of the clumps seen in \twelveCO{} and \thirthCO{} which enables the virial analysis of all clump populations.

\begin{figure*} 
\includegraphics[width=1.\textwidth]{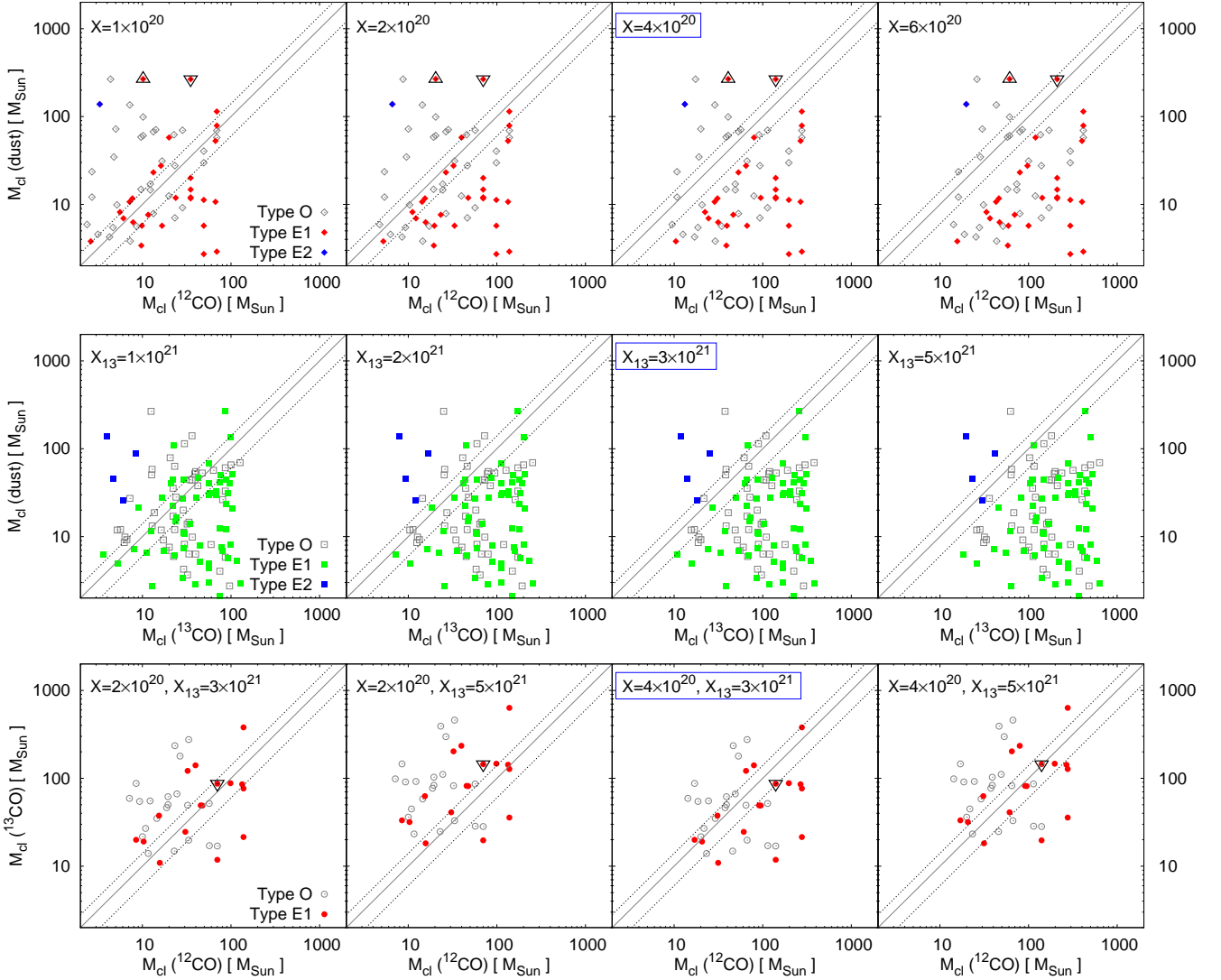}
\vspace{0.3cm}  
\caption{Comparison of clump masses from the associated pairs of Population 1 (absciss) and Population 2 (ordinate): \twelveCO-dust (top), \thirthCO-dust (middle) and \twelveCO-~\thirthCO(bottom). The test values of the $X$-factors are given in units [cm$^{-2}$K$^{-1}$km$^{-1}$s]; the adopted ones are put in a box. The dashed lines on both sides of the identity line represent the adopted uncertainty of mass estimates (see text). The triangles mark the dust cores identified with AFGL 961 and PL3 that were excluded from the analysis.}
\label{fig_mass-mass_diagram}
\end{figure*}

\subsection{Estimation of the {\it X}-factor}
\label{Estimation of X-factor}
Masses of condensations delineated in molecular-line maps are usually derived from object's intensity $W$ integrated over its area and line-of-sight through the factor $X=N(\mathrm{H}_2)/W$. Our first task in the analysis of the extracted clumps is to achieve a highly plausible estimate of $X$ for the tracers \twelveCO~and~\thirthCO{} based on clump associations. We use the masses of dust clumps  $M_{\rm cl}({\rm dust})$ as reference values for a comparison with the masses of their CO associates. The elaborated procedure of pixel-to-pixel determination of column density (Section \ref{Maps}) should yield $M_{\rm cl}({\rm dust})$ with uncertainties of better than 50\,\%
\citep{Roy_ea_13}.

Observational works on many nearby MCs indicate that the $X$-factor for \twelveCO~(hereafter, just $X$) in the Galaxy is approximately constant \citep{Dame_ea_01, BWL_13}. Theoretical models also show that a constant $X$ is justified, except in case of high metallicities \citep{Szucs_ea_16}. The widely adopted reference value for the Milky Way is: 
\begin{equation}
 \label{eq_X_12CO}
X_{\rm MW}(^{12}\mathrm{CO})\equiv X_{\rm MW}= 2\times10^{20}~\mathrm{cm}^{-2}\mathrm{K}^{-1}\mathrm{km}^{-1}\mathrm{s}
\end{equation}

However, numerical simulations including details of the hydrogen, carbon, and oxygen chemistry indicate a significant variation of the $X$-factor ($1\lesssim X/X_{\rm MW} \lesssim 10^{4}$) in the low-extinction regime $A_V\lesssim 3.5$ \citep{Glover_MacLow_11, Shetty_ea_11}. \citet{Ossenkopf_02} showed that $X$ might increase by two orders of magnitude from low- to high-density regimes due to optical depth effects. More than half of all associated \twelveCO~clumps in the RMC populate regions where such variations of $X$ should be expected. Therefore, instead of choosing a priori some average value of $X$ in RMC, we estimate it by comparing the masses of gaseous and dust clumps from the embedded pairs (type E). 

Our basic assumption is that, within the error bars, the mass of the embedded clump should be smaller than that of its counterpart in the other tracer: $M_1\ge M_2$ for pairs of Type E1 and $M_1\le M_2$ for pairs of Type E2 (cf. Fig. \ref{fig_criterion_clump_association}). Per definition, associated pairs of both types are separated by the identity line in the size-size diagram (see Fig. \ref{fig_appendix_R-R_diag} in the Appendix). Such a separation should also be observed in the mass-mass diagram if the Gaussian clumps in a pair are indeed spatially embedded.  

Under this assumption, we varied $X/X_{\rm MW}$ between $0.1$ and $10$ and calculate the masses of \twelveCO~clumps $M_{\rm cl}$\,(\twelveCO) for each test value. Then, by comparing $M_{\rm cl}$\,(\twelveCO) with $M_{\rm cl}$\,(dust) of the embedded dust associates, one can probe the range of plausible values of $X$ (Fig. \ref{fig_mass-mass_diagram}, top). Two pairs of clumps have to be excluded. The infrared sources AFGL 961 and PL3 (\citet{PhelpsLada1997}, clump D in \citealt{Poulton2008}) show irregular \twelveCO{} line profiles indicating strong self-absorption and outflows \citep[see][]{Schneider_ea_98} so that they hardly trace the column density structure. For completeness they are shown in Fig.~\ref{fig_mass-mass_diagram} but were excluded from further consideration. For all other clumps we adopt a possible 40\% uncertainty of the calculated masses of PPV clumps due to projection effects \citep{Beaumont_ea_13}.
Visual inspection of the mass-mass diagrams of the associated \twelveCO-dust pairs of type E1 (first row of Fig.~\ref{fig_mass-mass_diagram}) shows that within the error bars values of $X$ between $2\times10^{20}~\mathrm{cm}^{-2}\mathrm{K}^{-1}\mathrm{km}^{-1}\mathrm{s}$ and $6\times10^{20}~\mathrm{cm}^{-2}\mathrm{K}^{-1}\mathrm{km}^{-1}\mathrm{s}$ provide $M_{\rm cl}$(\twelveCO)$>M_{\rm cl}$\,(dust). 

The possible values of $X(^{13}\mathrm{CO})$ (hereafter, $X_{13}$) are obtained in analogous way. We probed the test range $1\le X_{13}/(8\times10^{20}~\mathrm{cm}^{-2}\mathrm{K}^{-1}\mathrm{km}^{-1}\mathrm{s}) \le 10$; some mass-mass diagrams are shown in the central row of Fig. \ref{fig_mass-mass_diagram}. The reference value corresponds to $N(\mathrm{H}_2)/N(^{13}\mathrm{CO})=7\times10^5$ \citep{FLW_82}, which is widely used in studies of large MC complexes \citep[e.g.][]{Nagahama_ea_98, ZXY_14}, combined with the LTE emissivity $W/N(^{13}\mathrm{CO})=9\times10^{-16}$~~K\,km\,s$^{-1}$\,cm$^2$ at a temperature of $T=15$~K \citep{Mangum_Shirley_15}. The objects of Types E1 and E2 form clearly separable groups on the mass-mass diagram. This enables us to constrain $X_{13}$ in a narrow range between $3\times 10^{21}~\mathrm{cm}^{-2}\mathrm{K}^{-1}\mathrm{km}^{-1}\mathrm{s}$ and $5\times 10^{21}~\mathrm{cm}^{-2}\mathrm{K}^{-1}\mathrm{km}^{-1}\mathrm{s}$ consistent with studies where the ratio $^{12}{\rm C}/^{13}{\rm C}$ has been probed \citep[e.g.][]{Blake_ea_87}. Due to the occurrence of both subtypes of embedding this constraint is much stronger than for \twelveCO. 

Finally we probe the associated clump pairs \thirthCO-\twelveCO~on the mass-mass diagrams using the limits on $X$ and $X_{13}$ factors obtained above (bottom row in Fig. \ref{fig_mass-mass_diagram}). Here it becomes clear that the constraint that all \thirthCO{} clumps are embedded in \twelveCO{} clumps, consequently having masses $M_{\rm cl}$\,(\thirthCO)$\le M_{\rm cl}$\,(\twelveCO), is only fulfilled if we use the combination of $X/X_{\rm MW}=2$ and $X_{13}=3\times 10^{21}~\mathrm{cm}^{-2}\mathrm{K}^{-1}\mathrm{km}^{-1}\mathrm{s}$. In this way, the concept of clump associations provides a very stringent limit on the $X$ and $X_{13}$ factors in the cloud if we assume that they are constant. The abovementioned estimates of $X/X_{\rm MW}$ and $X_{13}$ are taken as average values in the RMC to calculate clump masses.

\subsection{Size-mass relationships}
\label{Size-mass relationships}


\begin{figure*} 
\includegraphics[width=.7\textwidth]{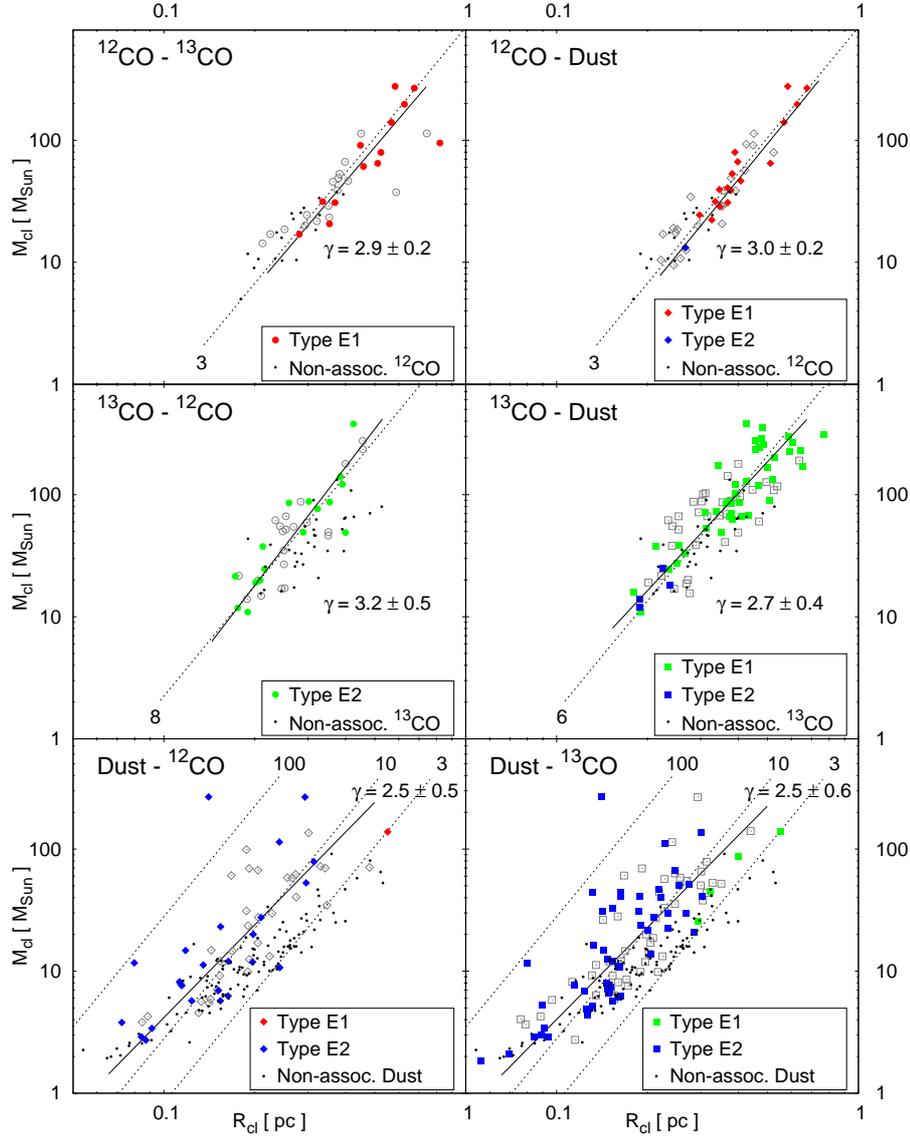}
\vspace{0.6cm}  
\caption{Size-mass diagrams of the clump populations. The symbols are the same as in Fig.~\ref{fig_mass-mass_diagram} but note the interchange of Types E1 and E2 depending on clumps of which tracer are taken to be Population 1. Dots show the corresponding locations of non-associated clumps of each tracer. The derived scaling relations (solid; cf. Table \ref{table_scaling_index_mass_Gauss}, column 5) and lines of constant mean density $\langle n\rangle$ in units $10^3$~\cc(dashed) are indicated.}
\label{fig_mass-size_Gaussian_assoc}
\end{figure*}

\begin{table}
\caption{Size-mass relationships for different clump populations.}
\label{table_scaling_index_mass_Gauss} 
\begin{center}
\begin{tabular}{c@{~~}c@{~~}c@{~\,}c@{~\,}c}
\hline 
\hline 
Tracer & Total population & \multicolumn{3}{c}{{\it Associated population}} \\ 
~ & $\gamma$ & Tracer & Clumps & $\gamma$ \\ 
\hline 
\twelveCO & $2.5 \pm{\scriptstyle 0.3}$ & \thirthCO & ~30 & $2.9 \pm{\scriptstyle 0.2}$ \\   
~ & ~ & Dust & ~40 & $3.0 \pm{\scriptstyle 0.2}$ \\ 
~ & ~ & \twelveCO & ~17 & $3.0 \pm{\scriptstyle 0.1}$ \\  
\hline
\thirthCO & $2.6 \pm{\scriptstyle 0.5}$ & \twelveCO & ~37 & $3.2 \pm{\scriptstyle 0.5}$ \\ 
~ & ~ & Dust & ~80 &$2.7 \pm{\scriptstyle 0.4}$ \\ 
~ & ~ & \thirthCO & ~22 & $2.9 \pm{\scriptstyle 0.2}$ \\ 
\hline
Dust & $2.3 \pm{\scriptstyle 0.5}$ & \twelveCO & ~58 & $2.5~{\scriptstyle \pm 0.5}$ \\ 
~ & ~ & \thirthCO & 105 & $2.5~{\scriptstyle \pm 0.6}$ \\ 
~ & ~ & Dust & ~13 & $2.7~{\scriptstyle \pm 0.7}$ \\
\hline 
\hline 
\end{tabular} 
\end{center}
\smallskip 
\end{table}

The mass-size relationship $M\propto R^{\gamma}$ reflects the scaling of density in molecular clouds \citep{Larson_81, Solomon_ea_87, Heyer_ea_09}. Table \ref{table_scaling_index_mass_Gauss} contains the mass scaling indices $\gamma$ obtained from the studied samples of Gaussian clumps in the RMC. The values for the full sample of clumps in each tracer (column 2) are about $\gamma=2.4$, consistent with other studies at scales below $1$~pc \citep[e.g.][]{Heithausen_ea_98, Hennebelle_Falgarone_12}. 

However, tightening the scope of consideration only to the {\it associated} clump populations yields an increase of the scaling indices $\gamma$ by 0.3--0.5. From the table it becomes clear that the associated clumps do not describe exactly the same material because it makes a difference whether we study e.g. the properties of CO clumps that have a dust counterpart or the properties of these dust counterparts.

For the associated CO clumps (Fig. \ref{fig_mass-size_Gaussian_assoc}, top and middle, and Table \ref{table_scaling_index_mass_Gauss}, column 5) $\gamma\simeq3$ within the uncertainties. In other words, these populations obey a relation of constant mean volume density $\langle n \rangle$ of the molecular gas in the range $3\times10^3$ to $8\times10^3$~\cc. The result is confirmed when associated clumps from a single tracer are considered (see \twelveCO-\twelveCO~and~\thirthCO-\thirthCO~in Table \ref{table_scaling_index_mass_Gauss}).
On the other hand, dust clumps with CO counterparts exhibit a shallower mass scaling that is only slightly higher than the one obtained from the full sample. Test runs of \Gauss~on the dust column-density map showed that the result is independent of the chosen noise level. 


\subsection{Cross-correlation between the maps}
\label{Cross-correlation between maps}

\begin{figure} 
\hspace{10mm}
\includegraphics[width=84mm]{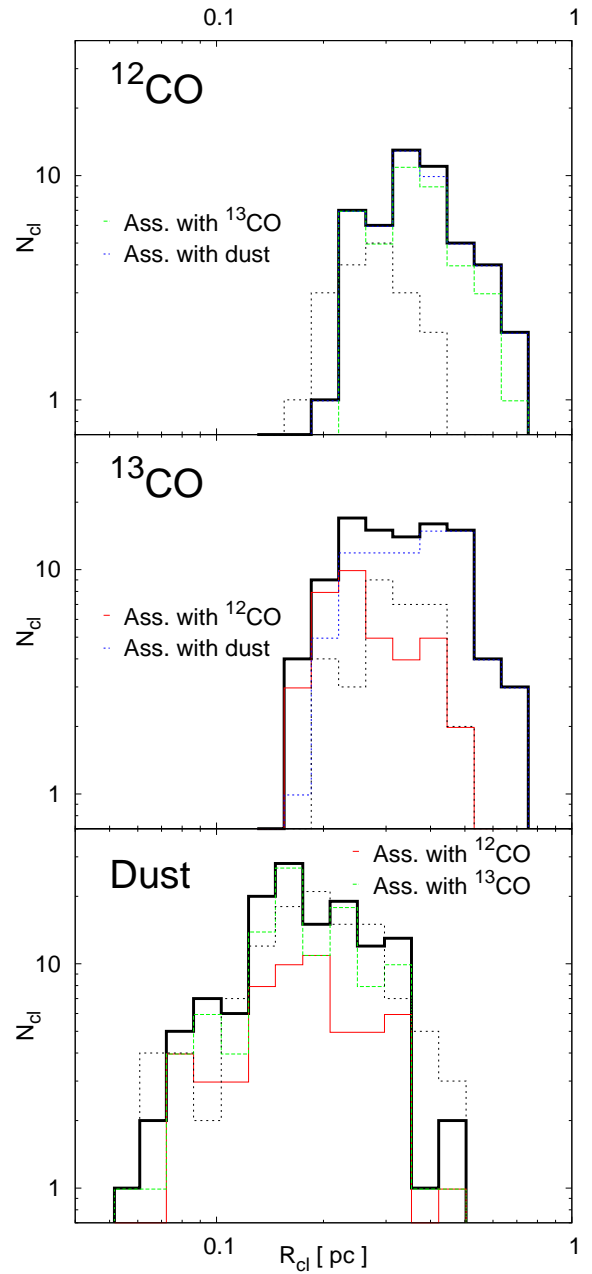}
\vspace{0.7cm}  
\caption{Distribution of clump sizes in the different populations. The solid lines show the distribution of associated clumps (black -- all, color -- with the specified tracer only), the dashed lines those of non-associated clumps. }
\label{fig_size-dist}
\end{figure}

The variation of the mass scaling index for associated clump populations prompt us to probe the strength of the association method. An appropriate way to do this is by study the cross-correlation between the maps in the corresponding tracers as a function of {\it abstract} spatial scale.

Comparing the size distributions of the associated and non-associated clumps in Figs.~\ref{fig_mass-size_Gaussian_assoc} and \ref{fig_size-dist} one can see that the non-associated \twelveCO{} clumps tend to be smaller by about a factor of two than those associated with dust or \thirthCO. For \thirthCO{} clumps the trend also holds for associations with \twelveCO{}, but not for associations with dust. In contrast there is no size dependence for associations from the viewpoint of dust clumps. Moreover, the size range and distribution of the \twelveCO{} clumps and the \thirthCO{} clumps is similar while the dust clumps are on average much smaller. Using a naive comparison of the size distributions of associated and non-associated clumps, one expects a better correlation of \twelveCO{} with the other tracers when increasing the scale. In contrast, from the dust clump size spectrum one would not expect a scale dependence of the correlation between dust and other tracers. When straightforwardly comparing \twelveCO{} and dust, both assumptions are obviously mutually exclusive.

The mutual size relations between different tracers can be directly measured by investigation of the cross-correlations between the three different maps as a function of the spatial scale. To this aim, we use the wavelet-based weighted cross-correlation (WWCC) tool \citep{ArshakianOssenkopf2016} that studies the degree of correlation of structures seen in a pair of maps as a function of the structure size. The method filters the two maps with a wavelet of a characteristic size so that only structures of that size remain visible and then computes the cross-correlation between the two filtered maps. In this way only structures of the same scale are compared. By using different wavelet sizes we obtain a spectrum of cross-correlation coefficients. Every point in the individual maps can be weighted by a noise weight to adjust the statistical significance of the result to the actual measurement uncertainties. In case of systematic shifts of characteristic structures between the two compared maps the WWCC can also measure their mutual displacement.
 
\begin{figure} 
\includegraphics[width=84mm]{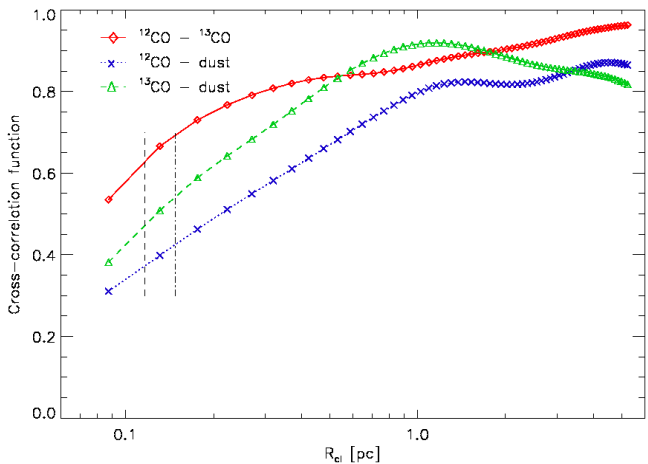}
\vspace{0.5cm}  
\caption{Spectra of wavelet-based cross-correlation coefficients as a function of the clump size scale for the three pairs of maps in the RMC. For a comparison to the clump size spectra, the wavelet scale $l$ has been divided by $8/\sqrt{8 \ln(2)}\approx 3.4$ to obtain corresponding clump radius units, $R_{\rm cl}$. The two vertical dashed lines indicate the resolution limit of the observations of 36$''$ (dust) and 46$''$ (\twelveCO{},\thirthCO{}).}
\label{fig_wwcc}
\end{figure}

Figure~\ref{fig_wwcc} shows the spectra of cross-correlation coefficients as a function of scale for all three pairs of maps of the RMC. The calibration of the wavelet spectrum for individual Gaussian clumps by \citet{ArshakianOssenkopf2016} enables translation of the wavelet scale $l$ into the radius of the clumps: $l\approx 3.4 R_{\rm cl}$. Thus the cross-correlation spectrum can be plotted in Fig.~\ref{fig_wwcc} directly on the clump radius scale and thus a direct comparison to Fig.~\ref{fig_mass-size_Gaussian_assoc} is possible. The spectra of the cross-correlation of \twelveCO{} with either of the other two maps show a monotonous increase towards larger scales. Individual small structures can significantly deviate between the different tracers, partially due to the noise impact at small scales, but towards larger scales all maps trace the same overall structure of the molecular cloud. For small and large clump sizes, the structures in both CO isotopologue maps show the best matches, providing the highest cross-correlation coefficient. At very large scales, the cross-correlation coefficient even approaches unity which suggests an extremely good match of the large-scale molecular distribution. In contrast, the cross-correlation with the dust shows a more complicated behaviour. The large dynamic range of the dust map traces both low-level extended material and small dense cores. The wide-spread low intensity emission also traced by \twelveCO{} hardly contributes to the clump extraction but dominates the cross-correlation function at large scales. Individual high-column-density clumps match the dense structures also traced by \thirthCO{}. On scales around 1~pc \thirthCO{} is even best correlated with the dust. Both species show the same prominent structures of that size. This indicates that for structure sizes around 1~pc the \thirthCO{} emission is a reasonable column density tracer, getting optically thicker at smaller scales so that it is suppressed relative to the dust below 1~pc and overamplified relative to the rest of the map at larger scales. Therefore, the correlation with dust drops again at large scales. \twelveCO{} is already optically thick at much larger scales. For that reason it shows a very extended emission and a good correlation with the wide-spread dust distribution at the large scales but a low correlation at small scales.

From the monotonously increasing cross-correlation between all maps at sizes below 1~pc we would expect to find more clump associations with increasing clump sizes. The effect of more \twelveCO{} associations with clump size is clearly visible in Figs.~\ref{fig_mass-size_Gaussian_assoc} and \ref{fig_size-dist}. At large $R_{\rm cl} > 0.4$~pc even all \twelveCO{} clumps are associated, suggesting a more perfect match than the cross-correlation shows.
However, looking from the perspective of the dust clumps, we find no size-dependence of the associated objects in Figs.~\ref{fig_mass-size_Gaussian_assoc} and \ref{fig_size-dist}. This is due to the tendency of the small and dense dust clumps to be embedded in \twelveCO{} and \thirthCO{} clumps as discussed in Sect.~\ref{Estimation of X-factor}. Because of the small size of the dust clumps those associations do not contribute to the scale-dependent cross-correlation coefficient. All extended dust emission does not affect the clump extraction procedure but contributes to the cross-correlation function.
The drop of the correlation between \thirthCO{} and dust at larger scales also cannot be traced by individual clumps due to the lack of very large clumps. In other words, the cross-correlation spectrum provides a comparison between the tracers which extends to larger scales than the clump associations.

To sum up the results from the WWCC test, clump associations seem to indicate only an overlap in physical space. Only if they occur for sizes with a high cross-correlation function, one can be sure that they also characterise approximately the same volume so that the density scaling of one species is transferable to the other one. Therefore we should always combine the approach described in Sect. \ref{Clump association} with the WWCC measure to exploit the strength of clump associations. The properties of associated populations can be always assessed from the two viewpoints of clumps within an associated pair, while the cross-correlation coefficient only provides a more general number, but at better spatial resolution and scale coverage.

\subsection{Virial analysis}
\label{Virial analysis}
Virial analysis is considered as a key to understand the star-forming properties and/or the evolutionary state of a MC. Below we assess the gravitational boundedness of the extracted \twelveCO{} and \thirthCO{} clumps in the RMC and that of their dust associates.

\subsubsection{Velocity dispersion}
\label{sect_size-linewidth}

\begin{figure*} 
\begin{center}
\includegraphics[width=.9\textwidth, keepaspectratio]{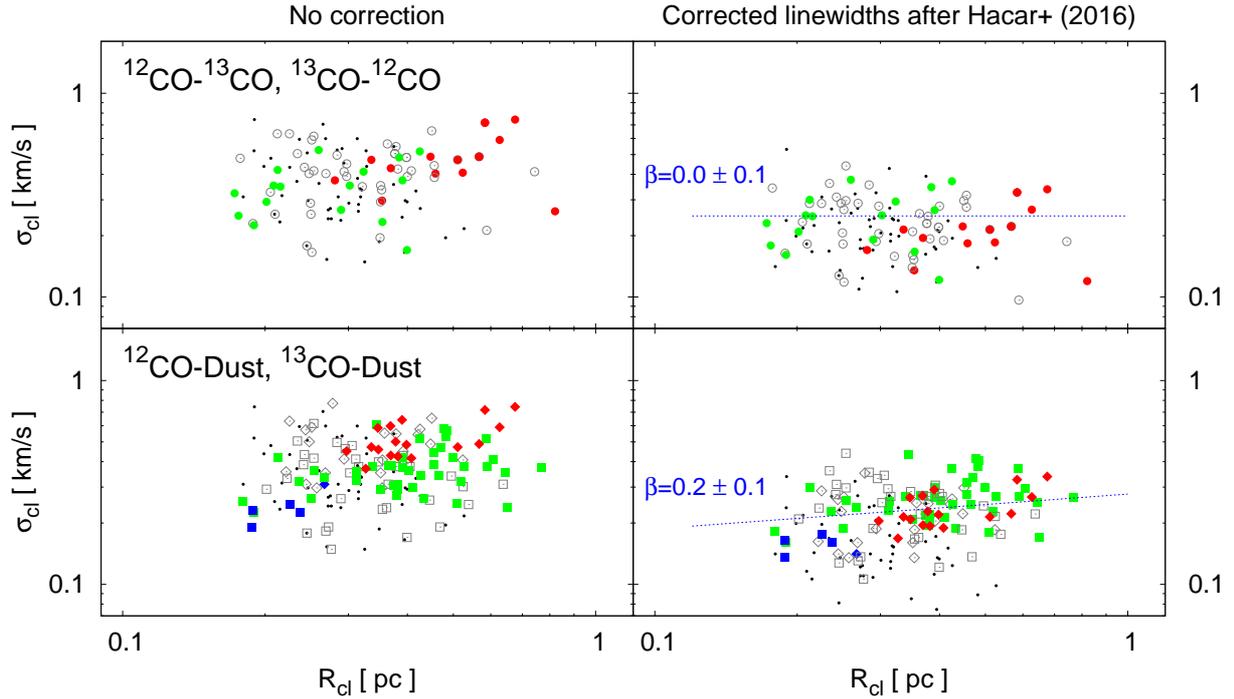}
\vspace{0.8cm}  
\caption{Diagram size vs. velocity dispersion for the clumps seen in \twelveCO{} or \thirthCO{}. The symbols are the same as in Fig.~\ref{fig_mass-size_Gaussian_assoc}. The upper plots show the associations of \twelveCO{} and \thirthCO{} clumps and the lower plots -- the associations of \twelveCO{}/\thirthCO{} with dust clumps. The left column represents the velocity dispersions calculated from the measured linewidths, the right column gives those after the optical depth correction following \citet{Hacar2016}, and the obtained slope for all associated (Types O+E) clumps. } 
\label{fig_sigma_scaling}
\end{center}
\end{figure*}

The virial parameter of a cloud or clump is defined as 
\begin{equation}
 \label{eq_virial_parameter}
  \alpha_{\rm vir}=\frac{5\sigma_{\rm cl}^2R_{\rm cl}}{GM_{\rm cl}}~, 
\end{equation}
with the velocity dispersion $\sigma_{\rm cl}$ and $G$ as the gravitational constant when assuming spherical geometry and constant density. It is often used as a tool to assess gravitational boundedness of objects -- those with $\alpha_{\rm vir}<2$ are considered bound \citep{McKee_Zweibel_92,BP_06}.

A critical parameter is the velocity dispersion $\sigma_{\rm cl}=\Delta v_{\rm cl}/(2\sqrt{2\ln2})$ that can be measured through the linewidth for all \thirthCO{} and \twelveCO{} clumps. Exploiting the strength of the clump associations we can also transfer the information on the velocity dispersion from the CO clumps to the dust clumps for which a direct measurement is impossible. The left column of Fig.~\ref{fig_sigma_scaling} shows the velocity dispersions of the associated \twelveCO{} and \thirthCO{} clumps vs. their sizes. We find an offset of the dispersion of the \twelveCO{} clumps relative to the \thirthCO{} clumps. In particular the \twelveCO{} clumps that are associated with dust have a higher velocity dispersion than the corresponding  \thirthCO{} clumps. The offset can be easily explained by optical depth broadening \citep[see e.g.][]{Phillips1979}.

A correction of the optical depth broadening can be performed following \citet{Hacar2016}. Those authors solved the radiative transfer equation for a Gaussian distribution of emitters in a homogeneous medium with optically thick conditions and derived a relation between the width of a Gaussian fit to an optically thick line and the width of the corresponding optically thin Gaussian profile as a function of the optical depth. In real clouds the velocity distribution may not be Gaussian, consisting of multiple velocity components along the line of sight. Nevertheless, their model matches our \Gauss~approach of decomposing the emission into Gaussian components that are treated separately. The test presented in Appendix \ref{appx_sigma_vel} confirms that we obtain a self-consistent picture when the correction after \citet{Hacar2016} is applied to the individual Gaussian velocity components. This correction is derived from their relation between the width of a Gaussian fit to an optically thick line and the width of the corresponding optically thin Gaussian profile as a function of the optical depth. A rough estimate of the optical depth can be obtained from the average $X$-factors obtained in Sect.~\ref{Estimation of X-factor}. The theoretical limit for the $X$-factor of the $1-0$ line of CO isotopes in optically thin LTE conditions is given by the energies and Einstein coefficients of the molecules \citep[see e.g][]{Mangum_Shirley_15}. Using the typical temperature in the RMC of 20~K \citep{Schneider_ea_10} we obtain an emissivity of $7\times10^{-16}$~K km.s$^{-1}/\mathrm{cm}^{-2}$ per column density of the emitter. Assuming a CO abundance $N({\rm CO})/N(\mathrm{H}_2)=2.4\times10^{-4}$ and an isotopic ratio $N({\rm CO})/N(^{13}\mathrm{CO})=70$ this predicts optically thin limits for the $X$-factors of $X=5\times10^{18}~\mathrm{cm}^{-2}\mathrm{K}^{-1}\mathrm{km}^{-1}\mathrm{s}$ and $X_{13}=4\times10^{20}~\mathrm{cm}^{-2}\mathrm{K}^{-1}\mathrm{km}^{-1}\mathrm{s}$. Assuming that the higher $X$ factors obtained in Sect.~\ref{Estimation of X-factor} are due to optically thick emission we find optical depths in the order of $100$ for \twelveCO{} and in the order of 5 for \thirthCO{}. Using the optical depth correction from \citet{Hacar2016} for these values indicates that the velocity dispersions of \twelveCO{} clumps are actually lower by a factor of about 2.2 compared to the measured ones, while for the \thirthCO{} clumps we should apply a correction factor of 1.4. Using these correction factors to the two linewidths we obtain a good match between the velocity dispersion measured through \twelveCO{} and \thirthCO{} (see Appendix~\ref{appx_sigma_vel}). The right column of Fig.~\ref{fig_sigma_scaling} shows the relation between clump size and velocity dispersions after applying the optical depth correction factors. The offset between the dispersion measured in \twelveCO{} and \thirthCO{} clumps is completely removed so that we can be confident to have a reliable measure of the actual clump velocity dispersion.

The resulting velocity dispersions cover only a small range with an average $0.25$~km s$^{-1}$ and standard deviation of $0.06$~km s$^{-1}$. The small clumps where all three tracers are associated have velocity dispersions of about $0.13$~km s$^{-1}$. Compared to the thermal value of 0.08~km s$^{-1}$ at 20~K, these are only slightly suprathermal. As there is no offset in the velocity dispersion between \twelveCO{} and \thirthCO{} clumps with and without association to dust clumps we do not expect any fundamental difference between them so that we can use the velocity dispersion measured in the molecular lines to assess the virial stability of all of the associated dust clumps.

The size-linewidth diagram of all associations of \twelveCO{} and \thirthCO{} clumps and those of CO clumps with dust counterparts (top right and bottom right panels of Fig. \ref{fig_sigma_scaling}, respectively) indicates a very weak relation. When fitting the velocity scaling relation $\sigma_{\rm cl}\propto R_{\rm cl}^{\beta}$ we obtain $\beta=0$ for the \twelveCO{}-\thirthCO{}associations. Interestingly, if one considers only the embedded clump populations (Type E) the scaling index $\beta$ increases up to 0.7, though the scatter is large -- see last column of Table \ref{table_beta_eps_CO}. This behaviour could be indicative of their evolutionary state as we comment in Sect. \ref{Discussion}. For clumps associated with dust the fit finds a weak size dependence of $\beta=0.2$. The lack of a velocity scaling relation for the \twelveCO{} and  \thirthCO{} clumps is in line with their constant density. If we assume that the region is dominated by a global large-scale velocity field and the turbulence dissipation is a function of the gas density all clumps with the same density should have about the same velocity dispersion.

\subsubsection{Relationship virial parameter vs. mass}
\label{Virial_analysis_M}

\begin{figure} 
\begin{center}
\hspace*{0.7cm}
\includegraphics[width=83mm]{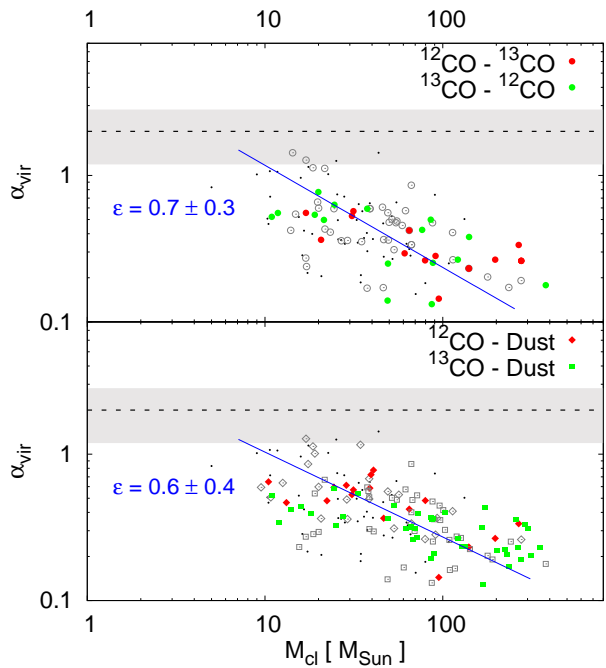}
\vspace{0.8cm}  
\caption{Virial analysis of the associated populations from diagram mass vs. virial parameter. The notation is the same like in Fig. \ref{fig_mass-size_Gaussian_assoc}. The slope for all associated clumps (Types O+E) is plotted (see text and Table \ref{table_beta_eps_CO}). Shaded area demarcates the effect of possible mass uncertainties on the assessment of clump boundedness ($\alpha_{\rm vir}\gtrless 2$). Non-associated \twelveCO{} and \thirthCO{} clumps (black dots) are shown for comparison.}
\label{fig_mass_alpha_vir}
\end{center}
\end{figure}

As evident from Fig. \ref{fig_mass_alpha_vir} and taking into account the assumed mass uncertainties, all associated CO clumps in the RMC are assessed as bound. The evolutionary state of the clumps can be characterised by the relation between virial parameter and mass, assumed to be a power law: $\alpha_{\rm vir}\propto M_{\rm cl}^{-\epsilon}$ \citep{BM_92, Dib_ea_07, Shetty_ea_10}. It is also recovered from combination of the mass-size relationship with power index $\gamma$ with a scaling relation of the velocity dispersion $\sigma_{\rm cl}\propto R_{\rm cl}^{\beta}$ which yields:
\begin{equation}
 \label{eq_relation_epsilon_beta_gamma}
 \epsilon=1-\frac{2\beta+1}{\gamma}~~.
\end{equation}

In Table \ref{table_beta_eps_CO} we give the slopes obtained by direct fitting of the samples in Fig. \ref{fig_mass_alpha_vir}. They are consistent with the values calculated from equation \ref{eq_relation_epsilon_beta_gamma}. Generally, the slope obtained for the embedded populations (Type E) seems to be shallower than the one when all associated clumps (Types O+E) are considered; while both fall within the range of estimates in star-forming regions sampled and explored by \citet[][see Table 2 there]{KPG_13}. We note that the obtained $\epsilon\sim0.6-0.7$ should not be interpreted as indicative for pressure-confined clumps, though similar to $\epsilon=2/3$ derived by \citet{BM_92} on this assumption. Virial parameters below unity point to substantial role of self gravity.  

\begin{table}
\caption{Scaling indices of virial parameter in respect to mass for the associated populations as derived directly from the diagram $M_{\rm cl}$ vs. $\alpha_{\rm vir}$ (column 3) and calculated from the scaling relations of mass and density (column 4). The slope $\beta$ of the relationship size vs. velocity dispersion is given for comparison.}
\label{table_beta_eps_CO} 
\begin{center}
\begin{tabular}{c@{~~}c@{~~}ccc}  
\hline 
\hline 
\multicolumn{5}{c}{\twelveCO-\thirthCO,~\thirthCO-\twelveCO} \\
~ & Type & $\epsilon$ & $\epsilon_{\rm calc}(\beta,\,\gamma)$ & $\beta$  \\ 
\hline 
~ & O + E & $0.7\pm{\scriptstyle 0.3}$ & $0.7\pm{\scriptstyle 0.2}$ & $0.0\pm{\scriptstyle 0.1}$ \\ 
~ & E & $0.3\pm{\scriptstyle 0.1}$ & $0.2\pm{\scriptstyle 0.2}$ & $~0.7\pm{\scriptstyle 0.3}$ \vspace*{3pt}\\
\hline
\multicolumn{5}{c}{\twelveCO{}-Dust,~\thirthCO{}-Dust} \\
~ & Type & $\epsilon$ & $\epsilon_{\rm calc}(\beta,\,\gamma)$ & $\beta$  \\ 
\hline 
~ & O + E & $0.6\pm{\scriptstyle 0.4}$ & $0.5\pm{\scriptstyle 0.1}$ & $0.2\pm{\scriptstyle 0.1}$ \\ 
~ & E & $0.3\pm{\scriptstyle 0.1}$ & $0.2\pm{\scriptstyle 0.2}$ & $0.7\pm{\scriptstyle 0.2}$  \\
\hline 
\hline 
\end{tabular} 
\end{center}
\smallskip 
\end{table}

\begin{figure} 
\hspace*{1cm}
\includegraphics[width=84mm]{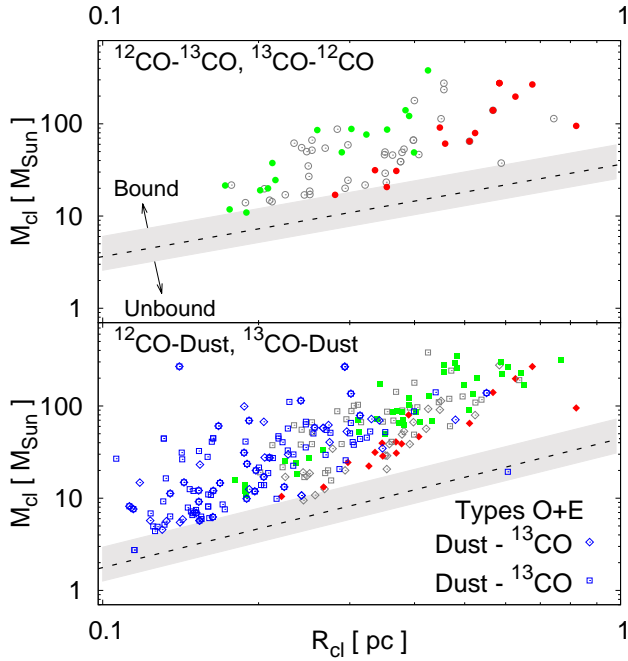}
\vspace{0.6cm}  
\caption{Virial analysis of the associated populations from mass-size diagrams. The notation of the CO populations (filled symbols for Type E, open - for Type O) is the same as in Fig. \ref{fig_mass-size_Gaussian_assoc}. Dashed line shows the virial mass $M_{\rm cl}(\alpha_{\rm vir}=2)$ as function of size, assuming velocity scalings as in Fig. \ref{fig_sigma_scaling}, right; shaded area denotes its uncertainty due to uncertainty of $\alpha_{\rm vir}$ (cf. Fig. \ref{fig_mass_alpha_vir}). }
\label{fig_M-R_vir}
\end{figure}

One may use the mass-size diagram as an additional tool to assess clump boundedness. From equation \ref{eq_virial_parameter} we define the virial mass of a clump as $M_{\rm cl,\,vir}\equiv M_{\rm cl}(\alpha_{\rm vir}=2)=5\sigma_{\rm cl}^2 R_{\rm cl}/2G$. Adopting scaling relation $\sigma_{\rm cl}\propto \sigma_0 (R_{\rm cl}/1~{\rm pc})^{\beta}$, the clump virial mass scales as $M_{\rm cl,\,vir}\propto R_{\rm cl}^{2\beta+1}$. Fig. \ref{fig_M-R_vir} displays the mass-size diagrams of associated CO clumps with plotted $M_{\rm cl,\,vir} (R_{\rm cl})$, assuming $\beta=0$ and taking $\sigma_0=0.25$~km s$^{-1}$ which is the average velocity dispersion for this population (cf. Fig. \ref{fig_sigma_scaling}). Apparently, the line $M_{\rm cl}=M_{\rm cl,\,vir}$ could serve as a lower mass limit of the bound clump population. Thus we use it to assess also the boundedness of dust clumps (Fig. \ref{fig_M-R_vir}, bottom) for which no velocity information is available. In that vein, the association of dust clumps with CO counterparts enables their stability analysis -- all but a few are classified as bound objects.

\section{Clump mass functions}	
\label{CMFs}
Clump mass functions (CMFs) are of particular interest because they could shed light on the origin of the stellar initial mass function (IMF). Below we present the derived CMFs in the RMC and probe their dependence on tracer, spatial association and the applied method to fit their high-mass parts. 

The high-mass end of the CMF is usually represented through a power-law function ${\rm d}\,N_{\rm cl}/{\rm d}\,\log M_{\rm cl}\propto M_{\rm cl}^{\Gamma}$ above some characteristic mass $M_{\rm ch}$ \citep[see e.g.][]{SG_90, Heithausen_ea_98,  Kramer_ea_98, Reid_ea_10, VDK_13}. However, the slope $\Gamma$ obtained from least-squares fit (LSF) is sensitive to the choice of bin size, sample size and $M_{\rm ch}$. In contrast, the maximum-likelihood (ML) fitting approach implemented in the method {\sc Plfit} by \citet{CSN_09} yields simultaneously $\Gamma$ and $M_{\rm ch}$ from the analysis of {\it unbinned} observational sample and provides a good precision for samples with $>50$ objects. The clump samples in the present study are rich enough (cf. Table \ref{table_clump_association_statistics}) to allow for the application of both methods. 
\begin{figure} 
\begin{center}
\includegraphics[width=88mm]{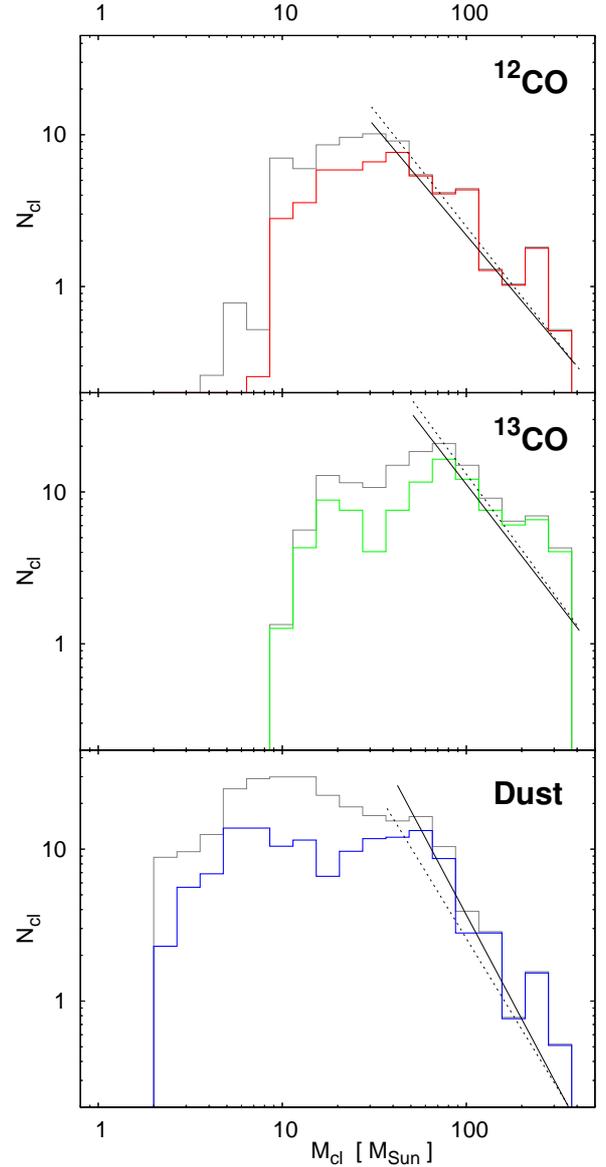}
\vspace{0.6cm}  
\caption{Mass functions of all clumps (grey) and associated clumps (colour) from a given tracer. The high-mass slopes derived through maximum-likelihood fitting for all clumps (dashed) and for the associated ones (solid) are plotted -- see Table \ref{table_mass_functions}.}
\label{fig_mass_functions}
\end{center}
\end{figure}

Information on the derived mass functions of the \twelveCO, \thirthCO~and dust clump populations is given in Table \ref{table_mass_functions} and is illustrated in Fig. \ref{fig_mass_functions}. In general, each distribution peaks close to the characteristic mass estimated from {\sc Plfit} and we adopted the latter as well in the LSF case. The non-associated clumps from all tracers contribute only to the low-mass part of the CMF. 

The obtained values of $\Gamma$ from both methods are similar while the ML fitting tends to yield steeper slopes. The slopes of the CO mass functions are close to the Salpeter IMF value $-1.3$ \citep{Sal55} -- also, when the full samples are considered. In contrast, the CMF of the dust clump population exhibits a substantially steeper slope. 

Several previous studies of the RMC included the derivation of the clump mass function -- in part by use of the same tracer and in a comparable mass range. All of them produced shallow slopes of a single power-law distribution, without a characteristic mass. \citet{WBS_95} applied the {\sc Clumpfind} algorithm to \twelveCO~and \thirthCO~maps and obtained a slope as shallow as $\Gamma=-0.27$ in mass range $10\lesssim M_{\rm cl}/M_\odot \lesssim4000$. They noticed, however, that only a few clumps from their sample are star-forming. \citet{Schneider_ea_98} derived $\Gamma=-0.6$ for Gaussian CO clump populations extracted both from KOSMA ($14\le M_{\rm cl}/M_\odot \le 1743$) and IRAM ($3\le M_{\rm cl}/M_\odot \le 50$) maps. Their virial analysis shows that a very small fraction of the objects are gravitationally bound. Extracting clumps from {\it Herschel} maps through the technique {\sc Getsources} and identifying them with young stellar sources from {\it Spitzer}, \citet{DiFran_ea_10} investigated separately samples of prestellar ($10\le M_{\rm cl}/M_\odot \le 300$) and starless ($2\le M_{\rm cl}/M_\odot \le 200$) cores. In both cases, a slope $\Gamma\simeq-0.8$ was obtained as the distribution peaks have been taken as characteristic masses and no assessment of the gravitational boundedness has been made. In terms of tracer and data source, our work on the dust population is comparable with that of \citet{DiFran_ea_10}. The much shallower $\Gamma$ found by those authors could be attributed to the applied different method for clump extraction and/or to the use of a single fit. If we adopt as a characteristic mass the mass-distribution peak for all dust clumps at $\sim10~M_\odot$ (cf. Fig. \ref{fig_mass_functions}, bottom), the corresponding slope is $\Gamma\simeq-0.7$ -- in agreement with the work of \citet{DiFran_ea_10}. Also, the results from both studies could reflect the lack of high-enough angular resolution to resolve clumps with sizes below $R_{\rm cl}\simeq0.2$~pc; cf. our Fig. \ref{fig_size-dist} (bottom) and Fig. 2 in \citet{DiFran_ea_10}. This size range roughly corresponds to masses $M_{\rm cl}\lesssim 20$~\Msol (Fig. \ref{fig_mass-size_Gaussian_assoc}).

\begin{table}
\caption{Parameters of clump mass functions from different tracers. Abbreviations of the fitting methods: ML - Maximum Likelihood, $D$ - Kolmogorov-Smirnov statistic of ML-fit goodness, LSF - weighted least-squares fit with logarithmic bin size $1.3$ and Poissonian error. The characteristic masses are taken from the ML output and given in solar units.}
\label{table_mass_functions} 
\begin{center}
\begin{tabular}{c@{~~}c@{~}c@{~}c@{~~}c@{~~~}c@{~}c@{~~}c@{~~~}c}
\hline 
\hline 
{\bf Tracer} & \multicolumn{4}{c|} {All clumps} & \multicolumn{4}{c} {All associated clumps} \\ \cline{2-9}
~ &  \multicolumn{3}{|c|} {ML} & \multicolumn{1}{|c|} {LSF} & \multicolumn{3}{|c|} {ML} & \multicolumn{1}{|c} {LSF} \\ \cline{2-9}
~ & $M_{\rm ch}$ & $D$ & $-\Gamma$ & $-\Gamma$ & $M_{\rm ch}$ & $D$ & $-\Gamma$ & $-\Gamma$ \\ 
\hline 
\twelveCO & $34$ & $0.09$ & $~1.5$ & $1.2\pm{\scriptstyle 0.1}$ & $34$ & $0.09$ & $1.4$ & $1.1 \pm{\scriptstyle 0.2}$ \\	
\thirthCO & $66$ & $0.10$ & $~1.6$ & $1.2\pm{\scriptstyle 0.1}$ & $66$ & $0.12$ & $1.6$ & $1.1\pm{\scriptstyle 0.2}$  \\
Dust & $44$ & $0.05$ & $~2.0$ & $1.5\pm{\scriptstyle 0.2}$ & $50$ & $0.07$ & $2.3$ & $2.3\pm{\scriptstyle 0.3}$ \\
\hline 
\hline 
\end{tabular} 
\end{center}
\end{table}

\section{Discussion}
\label{Discussion}


Possible caveats of the presented study are the uncertainties in the column-density maps and superposition effects. We find an average value for the $X$ factor that is somewhat higher than the Galactic average. Nevertheless, it gives realistic and self-consistent estimates of the masses of the embedded clump populations (Fig. \ref{fig_mass-mass_diagram}). We adopt also average uncertainties of size and mass estimates due superposition effects \citep{Beaumont_ea_13}. Individual uncertainty may vary with size; smaller clumps which are deeply embedded in cloud material are more susceptible to superposition. Objects of lower mean brightness could also misrepresent the real structure in the PPP space. However, such clumps do not affect the result on the mass scaling of the associated CO populations and, also, are not taken into account to derive the high-mass CMF.

Most molecular-line investigations of Galactic star-forming regions, by use of different clump-extraction techniques, result in mass scaling indices $\gamma$ close to $2$ or a few dex larger \citep[see][Table 1]{VDK_13}. This is reproduced from the derived mass-size relations in this work as far as {\it all} clumps from the considered population are taken into account (Table \ref{table_scaling_index_mass_Gauss}, column 2). The surprising result is the single steep mass scaling relation $\gamma\simeq3$ of the associated CO clumps, which corresponds to objects of constant mean volume density. How could one interpret this finding? Below we suggest a possible explanation from features of the CO detection and involving a model for clumps which are about to form stars. 

Critical density for excitation of CO molecules can explain the behaviour of associated clumps from these tracers. It gives us access to their nature that goes beyond the purely statistical result of a decomposition algorithm. The constant density of the CO clump associations is in line with the small dynamical range in densities that an individual molecular line is sensitive to \citep[see e.g.][]{Schneider2016c}. At lower densities the molecule is only subthermally excited, at higher densities it becomes quickly optically thick \citep{Draine_11, Klessen_Glover_15}. The density variation of the associated clumps probably stems from the different optical depths while both isotopes have about the same critical density for the collisional excitation. Hence, we can conclude that all associated CO clumps are real physical objects with at least the inferred density from Fig.~\ref{fig_mass-size_Gaussian_assoc}. In contrast, every non-biased clump decomposition mechanism tends to find clumps with constant column density \citep[see e.g.][]{Heyer_ea_09}. A fixed noise threshold and dynamic range of a map provide natural limits to the derived column density; an illustration for our clump samples is provided in the Appendix, Fig. \ref{fig_appendix_N_hist}. With {\sc Gaussclumps} larger structures can still be identified at average columns below the noise threshold of individual pixels so that we naturally expect a clump scaling exponent somewhat above two, in line with our values for the individual maps of $\gamma \approx 2.4$. The fact that associated dust clumps also show a steeper exponent indicates that the associations bias the dust clump sample to clumps with densities above the CO critical density, but -- as there is no upper limit for the measured column in the dust maps -- they can go towards higher densities. Most dust clumps can be regarded as the ``tips of the iceberg'' on top of the CO clumps, as seen through the types of embeddedness in Table \ref{table_associated_pairs_statistics} and Fig.~\ref{fig_mass-size_Gaussian_assoc}. Most dust clumps are embedded in CO clumps having significantly smaller radii. Hence, we can use the associations as a tool to identify significant physical entities above a certain density threshold.

The mean density of the associated CO clumps is about $3-8\times10^3$~\cc{} while dust emission as an optically thinner tracer allows for identification of their counterparts with $3\times10^3\lesssim \langle n\rangle\lesssim10^5$~\cc{} (Fig. \ref{fig_mass-size_Gaussian_assoc}, bottom). The shallower index $\gamma\sim2.5$ obtained for the sample of associated dust clumps, combined with the significantly larger scatter could be attributed to the lack of single density scaling law at scales below few $0.1$~pc -- the mass of structures of one and the same size could vary by orders of magnitude \citep{Falgarone_ea_92, Hennebelle_Falgarone_12}.

Another interesting result from our work is the virtual lack of a velocity-size relation for the associated CO clumps (Fig. \ref{fig_sigma_scaling}). It is consistent with the findings from a number of studies of clumps with similar size and with similar to higher densities in star-forming regions \citep{Casseli_Myers_95, Shirley_ea_03, Gibson_ea_09, Wu_ea_10}. We suggest an interpretation provided from the recent simulations of evolving MCs in a typical Galactic environment by \citet{Ibanez-Mejia_ea_16}. In their work, the identified structures of low density $\langle n\rangle\le5\times10^3$~\cc (i.e. traceable by CO emissions) display no dependence of the velocity dispersion with radius at early evolutionary phases, prior to the onset of self-gravity. This fact, along with the measured $0.35~{\rm km s}^{-1}\le\sigma\le0.6$~km s$^{-1}$, is explained with the formation of the considered objects through compression by converging flows. Their low velocity dispersions determine bound states: $\alpha_{\rm vir}$ less than $0.5$ and scaling with mass by $\epsilon\sim0.7$, in agreement with our results for all associated clumps in Sect. \ref{Virial_analysis_M}. On the other hand, further evolution of these clouds is affected strongly by self-gravity and leads to an increase of their velocity dispersions -- a linewidth-size relation appears, with index $\beta$ as high as $0.62\pm0.12$ (see Fig. 9 in \citealt{Ibanez-Mejia_ea_16}). Interestingly, such behaviour is found from virial analysis of our {\it embedded} clump populations (Table \ref{table_beta_eps_CO}). Whether this is indicative of their evolutionary state is speculative. A cross-identification with mapped young stellar sources and/or prestellar cores might shed light on this issue.   

The derived CMFs of the CO populations exhibit steep slopes in their high-mass parts, comparable to or larger than the one of the stellar IMF. Such slopes of $\Gamma$ are not exceptional -- similar values are derived from studies of other Galactic star-forming regions \citep[see review and the references in][]{VDK_13}, adopting  $M_{\rm ch}\gtrsim 10~M_\odot$ to separate intermediate-mass from high-mass part. In the cited work the observational CMFs were successfully modelled from statistical description of clump ensembles under the assumption of multi-scale, virial-like equipartition between gravitational and turbulent energy. In regard to this, we note again that the vast majority of the RMC clumps are found to be gravitationally bound by the performed virial analysis (Sect. \ref{Virial analysis}).

\section{Summary}
\label{Summary}

We study clump populations extracted from \twelveCO, \thirthCO~and \Herschel~dust-emission maps of the Rosette molecular cloud (RMC) using the clump-extraction algorithm \Gauss. By performing a cross-identification (association) between the populations from different tracers and subsequent physical analysis we can derive essential properties of the cloud structure:  
 \begin{itemize}

   \item Clump associations allow us to combine the information on the physical properties of the clumps that are measured by the different tracers. The assignment of the tracers to the same material is most reliable at scales of large cross-correlation coefficients, favouring the larger clumps in our study. By comparing clump masses we obtain a reliable estimate for the $X$ factors translating \twelveCO{} and \thirthCO{} intensities into column densities ($X=4\times10^{20}~\mathrm{cm}^{-2}\mathrm{K}^{-1}\mathrm{km}^{-1}\mathrm{s}$, $X_{13}=3\times 10^{21}~\mathrm{cm}^{-2}\mathrm{K}^{-1}\mathrm{km}^{-1}\mathrm{s}$). After correction for optical depth broadening we obtain a reliable measure for the velocity dispersion in the gaseous clumps that can be transferred to the dust clumps that have no separate velocity information.

   \item The associated Gaussian clumps extracted from CO tracers obey a single mass-size relation $M_{\rm cl}\propto R_{\rm cl}^3$ which implies approximately constant mean density about $3\times10^3$~\cc (\twelveCO) and  $6-8\times10^3$~\cc (\thirthCO). This behaviour can be explained by the small dynamical range in densities to which an individual molecular line is sensitive.
   
   \item The associated Gaussian clumps extracted from {\it Herschel} dust-emission map are usually embedded within the larger CO clumps representing their density peaks at scales a few tenths of pc where no single density scaling law (and no single mass-size relationship, respectively) is expected. This is reflected in a shallower mass scaling (slopes about $2.5$) than their CO associates. 
  
   \item All associated CO clumps (and all but a few of their dust counterparts) are assessed to be gravitationally bound and their location delineates the massive star-forming filaments and their junctions studied by \citet{Schneider_ea_12}. They display virtually no velocity-size relation, in consistence with the derived relation between their masses and virial parameters. We interpret this behaviour as indicative of low-density clumps formed through compression by converging flows and still not evolved under the influence of self-gravity. 
   
   \item The derived mass functions of the CO clump populations (associated or not) display a nearly Salpeter slope ($\Gamma\sim-1.3$), corresponding to the high-mass slope of the stellar initial mass function. In contrast, the mass functions of the dust clump populations are much steeper ($\Gamma\lesssim-2$).   

\end{itemize}

{\it Acknowledgement:} We are grateful to M. Heyer and J. Williams who graciously provided us with the FCRAO molecular-line maps of the Rosette region. We thank as well J. Ballesteros-Paredes for the helpful discussion on some results from this study and to the anonymous referee who's suggestions helped us to improve the readability of the paper.

This work was realised within the bilateral partnership agreement between the University of Sofia and the University of Cologne, part of the Priority program 1573 of the {\em Deutsche Forschungsgemeinschaft} (DFG). T.V. acknowledges partial support by the DFG under grant KL 1358/20-1 and V.O. and N.S. acknowledge support by the DFG under grant OS 177/2-2. R.S.K. thanks funding from the DFG in the Collaborative Research Center (SFB 881) "The Milky Way System" (subprojects B1, B2, and B8) and in the Priority Program SPP 1573 "Physics of the Interstellar Medium" (grant numbers KL 1358/18.1, KL 1358/19.2).

\label{lastpage}
\newpage
\appendix

\section{Characteristics of associated clump populations}
\label{Characteristics of associated pairs}

\begin{figure}  \hspace{10mm} 
\includegraphics[width=84mm]{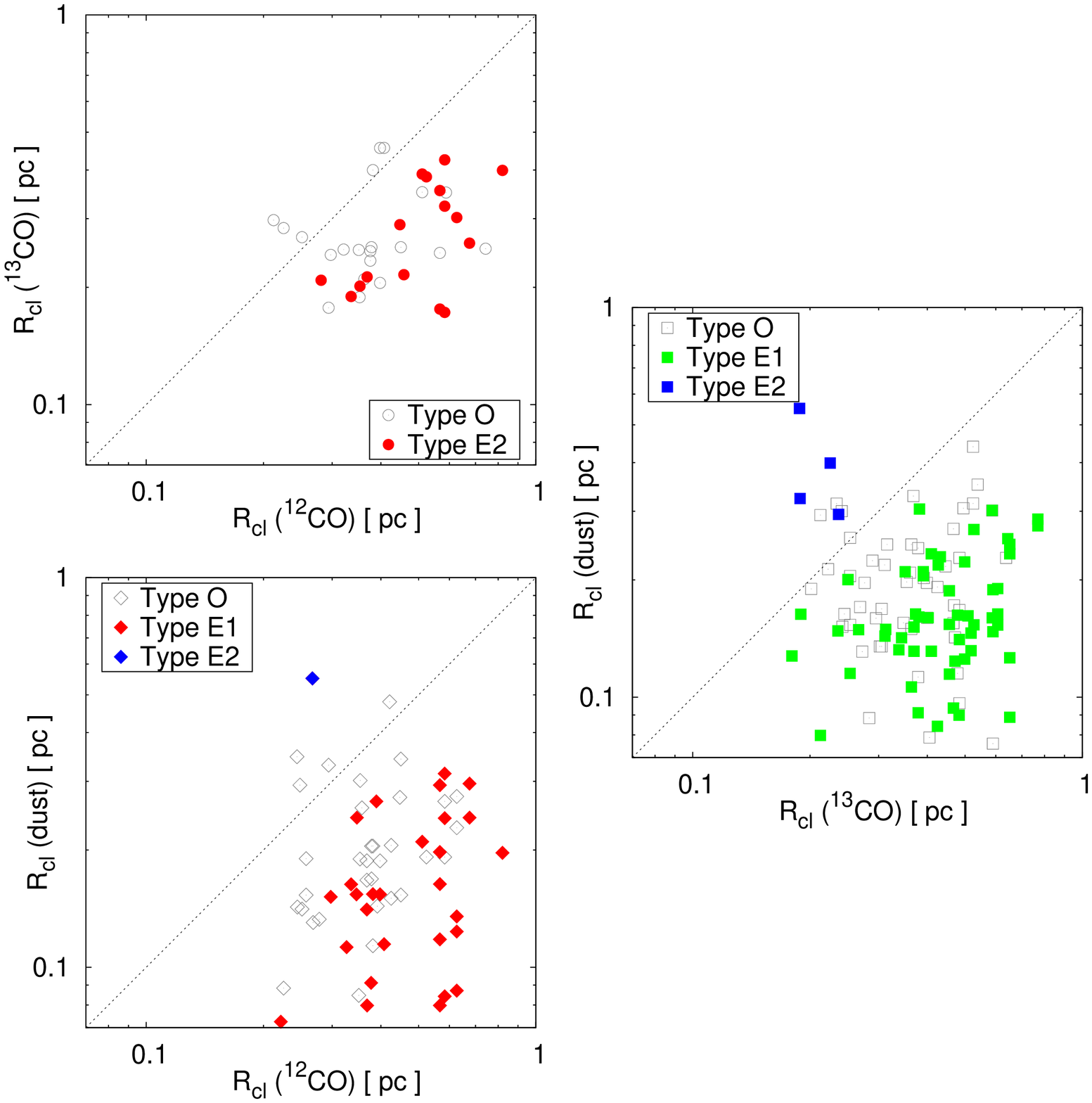}
\vspace{0.2cm}  
\caption{Sizes of associated Gaussian clump pairs, extracted from different tracers.}
\label{fig_appendix_R-R_diag}
\end{figure}

Tables \ref{table_12CO_assoc_clumps}, \ref{table_13CO_assoc_clumps} and \ref{table_Dust_assoc_clumps} provide full lists of the associated clumps and their physical parameters. The sizes of the associated objects from each pair of tracers are juxtaposed in Fig. \ref{fig_appendix_R-R_diag}. Embedded clumps (Types E1 or E2) are separated by the identity line per definition. In average, \twelveCO~clumps are about twice larger than their \thirthCO~embedded associates. The correlation of sizes of embedded \twelveCO/\thirthCO~-~dust pairs is weaker. 

\begin{figure}  \hspace{7mm} 
\includegraphics[width=74mm]{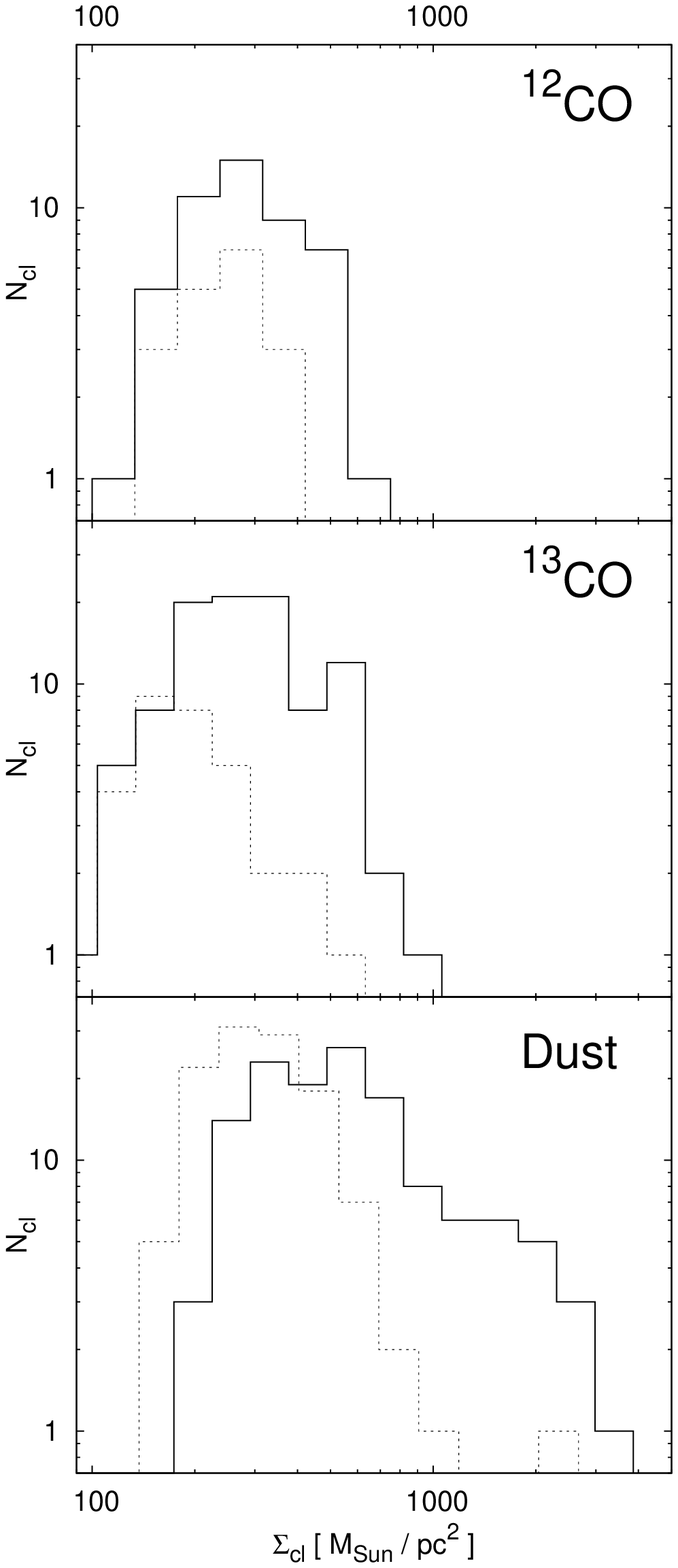}
\vspace{0.9cm}  
\caption{Distribution of mean surface density of clumps, extracted from different tracers. The solid lines show the distribution of clumps associated with any other tracer, the dashed lines those of non-associated clumps.}
\label{fig_appendix_N_hist}
\end{figure}

The distribution of mean surface densities for associated and non-associated clump populations from different tracers are plotted in Fig. \ref{fig_appendix_N_hist}.

\newpage
\begin{table*}
\caption{List of all associated \twelveCO~clumps.}
\label{table_12CO_assoc_clumps} 
\begin{center}
\begin{tabular}{r@{~~}c@{~~}c@{~~}c@{~~}c@{~~}c@{~~}crccc}
\hline 
\hline 
  \# & $l$ & $b$ & Axis 1 & Axis 2 & Orient. angle & Size & Mass & $\sigma$ & \thirthCO~ass. & Dust ass. \\
     & [deg] & [deg] & [\arcsec] & [\arcsec] & [deg] & [pc] & [\Msol] & [km s$^{-1}$] & ~ & ~ \\
\hline 
~~1 & 207.023926 & -1.820533 & ~74.6 & 109.7 & ~32.3 & 0.5833 & 276.5 & 1.69 & 1, 24, 80  &  10, 11, 34, 52, 282 \\
~~2 & 206.946793 & -1.811981 & ~81.7 & 115.5 & 153.1 & 0.6266 & 197.2 & 1.39 & 5   & 56, 78, 80, 177, 239 \\
~~4 & 207.323105 & -2.163783 & ~62.4 & 124.1 & ~64.5 & 0.5671 & 140.1 & 1.15 & 2, 116, 156 &  4, 26, 44, 117, 166 \\
~~7 & 206.992935 & -1.782242 & ~48.7 & ~78.3 & 165.2 & 0.3981 & ~66.6 & 1.14 & 3, 182 &  39, 286 \\
~~8 & 206.912140 & -1.869369 & ~56.0 & 117.8 & 147.0 & 0.5236 & ~79.5 & 0.96 & 56  & 187 \\
~10 & 206.945114 & -1.638992 & ~43.5 & 100.0 & ~66.3 & 0.4251 & ~92.9 & 1.36 & --  & 63, 224 \\
~11 & 207.027939 & -1.792006 & ~76.9 & ~47.5 & 170.8 & 0.3897 & ~79.8 & 1.51 & --  & 52 \\
~12 & 207.394547 & -1.941092 & ~55.3 & ~87.2 & ~55.7 & 0.4477 & ~91.0 & 1.15 & 41  & 38 \\
~16 & 206.905518 & -1.916839 & ~88.4 & ~57.2 & 174.0 & 0.4585 & ~61.0 & 0.95 & 179 & -- \\
~18 & 207.121262 & -1.881081 & ~39.0 & ~89.8 & ~30.4 & 0.3817 & ~53.2 & 1.00 & 11  & 14, 45, 163 \\
~22 & 206.951279 & -1.593222 & ~45.8 & 106.4 & 112.4 & 0.4502 & 113.5 & 1.54 & 85  & 57, 131 \\
~23 & 206.893539 & -1.799839 & ~36.4 & ~94.8 & 133.6 & 0.3785 & ~49.1 & 1.29 & 109 & 107, 148 \\
~26 & 206.794785 & -1.772153 & 177.1 & ~62.0 & ~~8.2 & 0.6756 & 267.4 & 1.75 & 23  & 71, 249 \\
~33 & 206.977036 & -1.837939 & ~81.4 & ~45.3 & 138.1 & 0.3917 & ~38.6 & 1.06 & --  & 60 \\
~34 & 207.425369 & -1.961631 & ~63.3 & ~63.2 & 173.4 & 0.4077 & ~46.5 & 0.98 & 9   & 92 \\
~36 & 207.049423 & -1.800294 & ~32.7 & 103.9 & ~58.6 & 0.3760 & ~38.9 & 1.19 & 18, 42 & -- \\
~37 & 206.923096 & -1.887656 & ~45.7 & ~65.8 & 128.3 & 0.3536 & ~20.7 & 0.70 & 124 & 257 \\
~42 & 207.093842 & -1.870303 & ~68.6 & ~43.2 & ~21.0 & 0.3511 & ~29.1 & 0.83 & 66  & 281 \\
~57 & 206.923035 & -1.846186 & ~32.9 & ~46.1 & 121.4 & 0.2510 & ~18.6 & 1.39 & 61  & 328 \\
~60 & 207.028488 & -1.838747 & ~29.5 & ~86.9 & ~39.5 & 0.3266 & ~22.3 & 0.87 & --  & 89 \\
~63 & 206.872894 & -1.840867 & ~40.4 & ~78.4 & ~72.8 & 0.3628 & ~45.8 & 1.33 & 302 & -- \\
~64 & 207.455322 & -1.968656 & ~43.2 & ~62.7 & ~30.9 & 0.3356 & ~31.4 & 1.11 & 251 & 293 \\
~66 & 207.154510 & -1.878117 & ~89.2 & ~33.8 & ~20.5 & 0.3541 & ~28.8 & 0.96 & --  & 15 \\
~74 & 207.305466 & -2.156178 & ~49.1 & ~66.7 & ~39.9 & 0.3689 & ~30.8 & 1.01 &  62 & 26 \\
~79 & 207.188660 & -1.926303 & ~49.9 & ~49.7 & 175.3 & 0.3211 & ~21.7 & 0.82 & 149 & -- \\
~81 & 207.000488 & -1.774714 & ~73.2 & ~39.4 & ~72.1 & 0.3463 & ~39.4 & 1.38 & --  & 286 \\
~91 & 207.272141 & -1.810972 & ~46.7 & ~69.9 & ~76.9 & 0.3683 & ~40.8 & 1.41 & --  & 1, 6, 7 \\
~96 & 207.314346 & -2.102122 & ~46.6 & ~94.7 & ~39.2 & 0.4284 & ~57.5 & 1.21 & --  & -- \\
102 & 207.078812 & -1.876608 & ~28.6 & ~55.6 & ~73.6 & 0.2572 & ~10.8 & 0.70 & --  & 65, 149 \\
103 & 207.066910 & -1.848575 & ~42.0 & ~28.1 & 152.2 & 0.2215 & ~10.4 & 0.84 & --  & 271 \\
107 & 207.429016 & -1.994303 & ~41.5 & ~41.4 & ~14.0 & 0.2670 & ~13.2 & 0.73 & --  & 66 \\
119 & 206.851364 & -1.798392 & ~37.0 & ~50.2 & 129.3 & 0.2779 & ~34.4 & 1.82 & --  & 236 \\
125 & 207.071335 & -1.782483 & ~70.4 & ~43.7 & ~13.4 & 0.3577 & ~38.6 & 1.30 & --  & 41 \\
128 & 207.156860 & -1.915303 & ~28.8 & ~66.1 & ~42.8 & 0.2810 & ~17.0 & 0.88 & 309 & -- \\
143 & 207.222549 & -1.637233 & ~34.3 & ~60.6 & 122.1 & 0.2938 & ~20.0 & 0.97 & 270 & 61 \\
151 & 207.228317 & -2.461564 & 152.8 & ~54.5 & 160.5 & 0.5882 & ~37.6 & 0.50 &  78 & -- \\
152 & 207.122803 & -1.639314 & ~65.2 & ~32.6 & 145.1 & 0.2975 & ~24.4 & 1.06 &  57 & 251 \\
158 & 206.854767 & -1.750297 & 109.9 & ~38.9 & ~57.5 & 0.4213 & ~56.6 & 1.28 & --  & 153 \\
160 & 207.183746 & -2.245544 & 243.2 & ~54.6 & 165.1 & 0.7431 & 114.0 & 0.97 & 154 & -- \\
173 & 207.311600 & -2.155100 & ~46.0 & ~32.2 & ~~1.9 & 0.2480 & ~17.5 & 1.18 & --  & 4 \\
190 & 207.219101 & -1.611764 & ~34.3 & ~31.6 & ~21.3 & 0.2123 & ~14.3 & 1.49 & 118 & -- \\
194 & 207.005981 & -1.927936 & ~37.9 & ~37.8 & 151.0 & 0.2442 & ~~9.5 & 0.73 & --  & 305 \\
200 & 207.349304 & -2.305289 & ~75.4 & 215.0 & ~82.9 & 0.8209 & ~95.0 & 0.62 & 98  & 201 \\
210 & 207.317719 & -1.900606 & ~42.9 & ~69.8 & ~43.3 & 0.3529 & ~23.3 & 0.79 & 148 & -- \\
218 & 206.910721 & -1.677203 & ~25.6 & ~55.8 & ~74.5 & 0.2437 & ~19.1 & 1.35 & --  & 219 \\
221 & 206.950302 & -1.560703 & ~32.1 & ~38.0 & 103.0 & 0.2252 & ~17.0 & 1.49 & 45  & 169 \\
227 & 206.799850 & -1.763731 & ~82.1 & ~35.3 & 143.4 & 0.3473 & ~28.7 & 1.08 & --  & 249 \\
262 & 207.676910 & -1.555356 & ~45.6 & 137.7 & 100.8 & 0.5109 & ~64.8 & 1.11 & 46, 258 & 69 \\
308 & 207.300903 & -2.127922 & ~44.3 & ~77.5 & ~49.2 & 0.3777 & ~38.9 & 1.18 & --  & 337 \\
376 & 207.123428 & -1.621158 & ~34.3 & ~50.5 & 151.8 & 0.2682 & ~12.6 & 0.83 & --  & 338 \\
\hline 
\hline 
\end{tabular} 
\end{center}
\smallskip 
\end{table*}

\newpage
\begin{table*}
\caption{List of all associated \thirthCO~clumps.}
\label{table_13CO_assoc_clumps} 
\begin{center}
\begin{tabular}{r@{~~}c@{~~}c@{~~}c@{~~}c@{~~}c@{~~}crccc}
\hline 
\hline 
  \# & $l$ & $b$ & Axis 1 & Axis 2 & Orient. angle & Size & Mass & $\sigma$ & \twelveCO~ass. & Dust ass. \\
     & [deg] & [deg] & [\arcsec] & [\arcsec] & [deg] & [pc] & [\Msol] & [km s$^{-1}$] & ~ & ~ \\
\hline 
~~1 & 207.017502 & -1.815992 & ~40.92 & 106.09 & ~44.1 & 0.4248 & 380.0 & 1.22 & 1 & 10, 282 \\
~~2 & 207.322815 & -2.167897 & ~38.79 & ~77.76 & ~47.5 & 0.3542 & ~86.4 & 0.55 & 4 &  117    \\
~~3 & 206.988449 & -1.784817 & 127.71 & ~39.13 & 171.2 & 0.4558 & 274.6 & 1.04 & 7 & 39, 286 \\
~~5 & 206.939468 & -1.809586 & ~51.31 & ~42.77 & ~~8.0 & 0.3021 & ~87.6 & 0.83 & 2 & 56      \\
~~6 & 207.127731 & -1.835806 & ~89.29 & ~62.76 & ~34.2 & 0.4827 & 350.1 & 1.25 & -- & 211    \\
~~7 & 207.284317 & -2.148969 & ~68.38 & ~47.76 & ~~3.9 & 0.3685 & 141.5 & 0.90 & -- & 111    \\
~~8 & 207.560471 & -1.713642 & ~86.50 & ~61.72 & 165.6 & 0.4711 & 240.2 & 1.10 & -- &  25, 137\\
~~9 & 207.423874 & -1.951617 & 135.91 & ~36.78 & ~23.7 & 0.4559 & 234.1 & 0.91 & 34 & 92      \\
~11 & 207.128937 & -1.875972 & ~85.09 & ~45.17 & ~11.4 & 0.3998 & 178.0 & 0.92 & 18 & 319     \\
~15 & 207.294266 & -2.092403 & ~43.07 & ~54.01 & ~29.6 & 0.3110 & 102.5 & 0.97 & -- & 40      \\
~18 & 207.045258 & -1.817914 & ~31.35 & ~42.06 & 122.7 & 0.2341 & ~61.5 & 1.19 & 36 & 34      \\
~19 & 207.259964 & -1.812581 & ~60.77 & ~92.63 & 141.2 & 0.4838 & 257.6 & 1.34 & -- &  1, 7, 64, 242 \\
~20 & 206.996277 & -1.816097 & ~44.21 & ~52.78 & ~87.8 & 0.3114 & ~70.9 & 0.76 & -- & 60  \\
~21 & 206.870819 & -1.874375 & ~43.96 & ~64.56 & ~84.6 & 0.3435 & 172.7 & 1.43 & -- & 20  \\
~23 & 206.797562 & -1.770117 & ~33.51 & ~48.37 & ~25.3 & 0.2596 & ~85.2 & 1.24 & 26 & --  \\
~24 & 207.012497 & -1.822536 & ~23.50 & ~30.41 & ~43.5 & 0.1724 & ~21.3 & 0.76 & 1 & --   \\
~25 & 206.908569 & -1.784900 & ~55.99 & ~40.10 & 168.4 & 0.3055 & 100.2 & 1.13 & -- & 222 \\
~28 & 207.067215 & -1.814750 & ~63.87 & ~86.20 & 147.3 & 0.4784 & 287.7 & 1.37 & -- & 22  \\
~30 & 207.290100 & -2.188197 & ~92.71 & ~47.29 & ~~7.0 & 0.4269 & 128.0 & 0.81 & -- & 43  \\
~34 & 207.310562 & -1.822886 & 115.00 & ~72.71 & ~12.1 & 0.5896 & 223.3 & 0.89 & -- &  23, 101, 160, 364 \\
~35 & 207.307831 & -2.132703 & ~41.21 & ~83.83 & 105.9 & 0.3790 & ~85.1 & 0.64 & -- & 337          \\
~37 & 207.154877 & -1.880883 & ~81.50 & 101.70 & 160.5 & 0.5871 & 298.8 & 1.22 & -- &  15, 51, 319 \\
~41 & 207.390228 & -1.946653 & ~39.24 & ~51.37 & ~25.7 & 0.2895 & ~48.9 & 0.63 & 12  & --          \\
~42 & 207.036850 & -1.788664 & ~27.32 & ~54.04 & ~28.0 & 0.2477 & ~49.8 & 0.95 & 36  & --          \\
~44 & 207.316818 & -1.877236 & ~31.34 & ~76.33 & ~30.6 & 0.3154 & ~66.3 & 0.89 & --  & 67          \\
~45 & 206.949249 & -1.564369 & ~30.21 & ~64.12 & ~87.8 & 0.2838 & ~87.1 & 1.17 & 221 & 169         \\
~46 & 207.676697 & -1.557078 & ~38.90 & ~94.31 & ~94.8 & 0.3905 & 121.0 & 0.88 & 262 & 69          \\
~47 & 207.589127 & -1.758081 & ~38.83 & ~82.22 & ~55.0 & 0.3643 & ~85.1 & 0.90 & --  & 12          \\
~48 & 207.144730 & -1.821492 & ~59.10 & ~88.71 & 121.4 & 0.4669 & 119.2 & 0.81 & --  & 47          \\
~49 & 207.082047 & -1.866792 & ~76.03 & 116.36 & 111.3 & 0.6065 & 265.1 & 0.97 & --  &  19, 65, 149\\
~50 & 207.210541 & -1.835375 & ~46.60 & ~59.03 & 158.2 & 0.3382 & ~72.6 & 0.89 & --  & 46          \\
~51 & 207.286285 & -2.424056 & ~59.84 & ~79.79 & 130.1 & 0.4456 & 109.3 & 0.61 & --  & 198         \\
~52 & 207.183502 & -1.783253 & ~49.66 & ~66.20 & 145.3 & 0.3697 & ~88.0 & 0.90 & --  & 9, 183, 358 \\
~54 & 207.782776 & -1.784194 & ~63.08 & 110.68 & ~87.7 & 0.5388 & 116.6 & 0.58 & --  & 104         \\
~56 & 206.905136 & -1.871392 & ~39.42 & ~90.14 & 141.8 & 0.3843 & 140.0 & 1.14 & 8   & --          \\
~57 & 207.108276 & -1.642528 & ~26.91 & ~52.42 & 116.9 & 0.2422 & ~55.1 & 1.02 & 152 & 251         \\
~59 & 207.172119 & -1.845550 & ~54.06 & ~73.01 & ~23.2 & 0.4051 & ~96.3 & 0.89 & --  & 194         \\
~61 & 206.922104 & -1.841356 & ~32.72 & ~53.15 & ~62.2 & 0.2689 & ~54.4 & 0.95 & 57  & --          \\
~62 & 207.304169 & -2.155556 & ~45.78 & ~23.75 & ~15.4 & 0.2126 & ~37.5 & 0.99 & 74  & 4, 26       \\
~65 & 207.105698 & -1.808583 & ~35.70 & ~42.22 & ~93.6 & 0.2503 & ~27.3 & 0.62 & --  & 98          \\
~66 & 207.089584 & -1.866308 & ~47.54 & ~31.48 & ~~1.8 & 0.2494 & ~34.8 & 0.68 & 42  & --          \\
~69 & 207.304062 & -2.532933 & 114.89 & ~58.13 & ~~9.2 & 0.5270 & 201.5 & 0.89 & --  & 21, 128     \\
~71 & 207.344818 & -1.948906 & ~23.89 & ~32.51 & ~59.6 & 0.1797 & ~15.9 & 0.60 & --  & 243         \\
~72 & 207.360107 & -1.945022 & ~26.36 & ~46.31 & ~71.0 & 0.2253 & ~24.9 & 0.58 & --  & 84          \\
~78 & 207.243835 & -2.449700 & ~35.08 & ~84.04 & 149.0 & 0.3501 & ~46.0 & 0.46 & 151 & --          \\
~79 & 207.277222 & -1.864947 & ~45.17 & 142.92 & ~11.5 & 0.5181 & 133.1 & 0.75 & --  & 208, 366    \\
~80 & 207.027939 & -1.828756 & ~63.65 & ~39.44 & 173.5 & 0.3230 & ~76.2 & 0.97 & 1   & --          \\
~81 & 207.317184 & -1.838792 & ~64.89 & ~49.17 & ~20.0 & 0.3642 & ~69.2 & 0.73 & --  & 24, 109     \\
~85 & 206.935623 & -1.601728 & ~26.51 & ~58.29 & 122.8 & 0.2535 & ~51.5 & 0.98 & 22  & 131         \\
~87 & 207.826660 & -1.849742 & ~72.84 & 140.08 & 140.4 & 0.6513 & 168.1 & 0.56 & --  & 29, 95, 195, 318\\
~91 & 207.367767 & -1.849053 & ~56.38 & 110.39 & ~53.2 & 0.5087 & ~89.3 & 0.59 & --  & 171          \\
~92 & 207.206085 & -1.866494 & 103.14 & ~33.78 & 160.4 & 0.3806 & ~69.6 & 0.73 & --  & 336          \\
~93 & 207.586731 & -1.727814 & ~65.47 & ~59.35 & ~23.7 & 0.4020 & ~86.3 & 0.85 & --  & 139          \\
~94 & 207.700165 & -1.920394 & 152.63 & ~65.29 & 179.6 & 0.6437 & 228.4 & 0.83 & --  & 49           \\
~96 & 207.361786 & -1.904942 & ~49.50 & ~74.21 & ~30.9 & 0.3908 & 102.3 & 0.99 & --  & 180          \\
~98 & 207.343964 & -2.297911 & ~66.36 & ~57.77 & ~18.6 & 0.3992 & ~48.8 & 0.40 & 200 & 201          \\
~99 & 207.299927 & -1.801072 & ~57.88 & 103.53 & 161.3 & 0.4991 & 166.2 & 0.99 & --  & 48, 261, 294 \\
100 & 207.554169 & -1.753769 & ~52.07 & ~70.92 & ~69.6 & 0.3918 & ~95.6 & 1.01 & --  & 168          \\
103 & 207.330521 & -2.350303 & ~44.85 & ~44.88 & 123.2 & 0.2893 & ~38.5 & 0.62 & --  & 241          \\
104 & 207.072906 & -1.796483 & ~27.39 & ~56.46 & ~92.9 & 0.2535 & ~38.4 & 0.85 & --  & 41, 360      \\
106 & 207.788742 & -1.878017 & 104.28 & 136.30 & ~73.8 & 0.7687 & 311.4 & 0.88 & --  & 59, 205      \\
\hline 
\hline 
\end{tabular} 
\end{center}
\end{table*}

\begin{table*}
\contcaption{~} 
\label{table_13CO_assoc_clumps_continued} 
\begin{center}
\begin{tabular}{r@{~~}c@{~~}c@{~~}c@{~~}c@{~~}c@{~~}crccc}
\hline 
\hline 
  \# & $l$ & $b$ & Axis 1 & Axis 2 & Orient. angle & Size & Mass & $\sigma$ & \twelveCO~ass. & Dust ass. \\
     & [deg] & [deg] & [\arcsec] & [\arcsec] & [deg] & [pc] & [\Msol] & [km s$^{-1}$] & ~ & ~ \\
\hline 
109 & 206.884689 & -1.796822 & ~28.75 & ~53.76 & 122.8 & 0.2535 & ~66.3 & 1.45 & 23  & 148          \\
111 & 207.572250 & -1.711033 & 103.15 & ~43.55 & 159.0 & 0.4322 & ~67.8 & 0.62 & --  & 8            \\
114 & 207.266602 & -1.772617 & ~33.14 & ~51.60 & ~80.3 & 0.2666 & ~33.1 & 0.79 & --  & 346          \\
116 & 207.311188 & -2.183314 & ~23.50 & ~61.47 & ~94.8 & 0.2451 & ~16.9 & 0.42 & 4   & 166          \\
118 & 207.232056 & -1.610922 & ~35.40 & ~60.12 & 122.0 & 0.2975 & ~58.8 & 0.92 & 190 & --           \\
122 & 207.020020 & -1.841539 & ~45.00 & ~76.75 & 115.8 & 0.3789 & ~89.0 & 0.89 & --  & 11, 89       \\
124 & 206.922455 & -1.882825 & ~27.41 & ~35.62 & 122.8 & 0.2014 & ~18.9 & 0.69 & 37  & 257          \\
126 & 207.231674 & -1.746394 & ~62.60 & ~37.58 & 173.2 & 0.3127 & ~52.7 & 0.84 & --  & 260          \\
127 & 207.433212 & -1.983422 & ~34.11 & ~24.90 & 179.7 & 0.1879 & ~11.9 & 0.45 & --  & 66           \\
129 & 207.083786 & -1.851375 & ~72.79 & ~46.01 & ~~6.2 & 0.3732 & ~65.6 & 0.72 & --  & 19           \\
134 & 207.366562 & -1.783275 & ~74.09 & ~39.19 & 157.3 & 0.3475 & ~67.2 & 0.94 & --  & 306          \\
140 & 206.931046 & -1.628206 & ~33.37 & ~62.96 & ~58.4 & 0.2955 & ~71.4 & 1.09 & --  & 33           \\
143 & 207.399872 & -1.378078 & ~48.45 & 121.76 & 131.3 & 0.4953 & 126.0 & 0.80 & --  & 133          \\
148 & 207.327835 & -1.905264 & ~28.33 & ~30.20 & ~20.3 & 0.1886 & ~14.0 & 0.54 & 210 & 108          \\
149 & 207.193726 & -1.922300 & ~23.51 & ~63.96 & ~28.2 & 0.2500 & ~26.8 & 0.60 & 79  & --           \\
150 & 207.891022 & -1.810161 & 111.83 & ~47.21 & 171.7 & 0.4685 & ~60.0 & 0.45 & --  & 273          \\
152 & 207.243652 & -1.566853 & ~26.01 & ~53.79 & 119.0 & 0.2412 & ~38.0 & 0.91 & --  & 77           \\
154 & 207.145691 & -2.245997 & ~25.68 & ~59.11 & 139.2 & 0.2512 & ~17.0 & 0.39 & 160 & --           \\
156 & 207.332291 & -2.166375 & ~23.71 & ~31.35 & ~68.0 & 0.1758 & ~11.8 & 0.59 & 4   & --           \\
158 & 207.645599 & -1.899469 & ~56.06 & ~55.99 & 132.7 & 0.3613 & ~40.7 & 0.56 & --  & 130          \\
165 & 207.762634 & -1.927164 & ~74.98 & ~88.25 & ~71.3 & 0.5245 & 108.4 & 0.71 & --  & 42, 97       \\
175 & 207.170944 & -1.798603 & ~46.96 & ~85.99 & 126.6 & 0.4097 & ~66.3 & 0.70 & --  & 134          \\
179 & 206.901352 & -1.911522 & ~39.46 & ~28.30 & ~39.6 & 0.2155 & ~24.5 & 0.82 & 16  & --           \\
182 & 206.989166 & -1.772997 & ~36.02 & ~28.11 & 169.0 & 0.2052 & ~19.8 & 0.77 & 7   & --           \\
189 & 207.681931 & -1.643536 & ~34.10 & ~52.33 & ~67.3 & 0.2724 & ~20.0 & 0.45 & --  & 124          \\
202 & 207.339600 & -2.567433 & ~45.69 & ~76.82 & ~64.3 & 0.3820 & ~62.2 & 0.69 & --  & 120          \\
203 & 207.388718 & -1.194906 & 212.27 & ~45.99 & ~38.8 & 0.6371 & 189.6 & 0.73 & --  & 28           \\
216 & 207.482956 & -1.700631 & ~38.51 & ~45.15 & ~83.3 & 0.2689 & ~18.6 & 0.43 & --  & 259          \\
223 & 207.294281 & -1.751072 & ~23.79 & ~56.71 & ~94.4 & 0.2368 & ~18.1 & 0.53 & --  & 265          \\
248 & 207.376724 & -2.361628 & ~34.79 & ~38.45 & 105.8 & 0.2358 & ~24.2 & 0.75 & --  & 207          \\
251 & 207.455704 & -1.967514 & ~36.34 & ~23.74 & ~31.7 & 0.1894 & ~10.9 & 0.53 & 64  & 293          \\
258 & 207.668015 & -1.555822 & ~30.50 & ~97.12 & ~91.2 & 0.3509 & ~49.3 & 0.69 & 262 & 69           \\
262 & 206.883255 & -1.889175 & ~26.83 & ~44.40 & 119.3 & 0.2225 & ~21.6 & 0.78 & --  & 74           \\
270 & 207.211227 & -1.640056 & ~31.82 & ~23.74 & 165.6 & 0.1772 & ~21.6 & 1.13 & 143 & --           \\
275 & 207.321793 & -2.293533 & ~29.75 & ~61.60 & 140.5 & 0.2761 & ~15.5 & 0.35 & --  & 201          \\
302 & 206.871368 & -1.844628 & ~23.69 & ~44.76 & ~73.5 & 0.2100 & ~14.8 & 0.60 & 63  & --           \\
309 & 207.154953 & -1.910250 & ~24.03 & ~43.51 & ~37.9 & 0.2085 & ~19.9 & 0.83 & 128 & --           \\
\hline 
\hline 
\end{tabular} 
\end{center}
\end{table*}

\newpage
\begin{table*}
\caption{List of all associated dust clumps.}
\label{table_Dust_assoc_clumps} 
\begin{center}
\begin{tabular}{r@{~~}c@{~~}c@{~~}c@{~~}c@{~~}c@{~~}crcc}
\hline 
\hline 
  \# & $l$ & $b$ & Axis 1 & Axis 2 & Orient. angle & Size & Mass & \twelveCO~ass. & \thirthCO~ass. \\
     & [deg] & [deg] & [\arcsec] & [\arcsec] & [deg] & [pc] & [\Msol] & ~ & ~ \\
\hline 
  ~~1 & 207.263519 & -1.812227 & 22.850 & 20.799 &  141.7 & 0.1405 & 267.7 & 91 & 19 \\ 
  ~~4 & 207.316498 & -2.152799 & 41.092 & 50.378 &  ~87.0 & 0.2933 & 266.4 & 4, 173 & 62 \\
  ~~6 & 207.281265 & -1.803322 & 21.302 & 39.748 &  123.1 & 0.1876 & ~99.0 & 91 & -- \\
  ~~7 & 207.274597 & -1.824577 & 25.311 & 26.595 &  ~74.0 & 0.1672 & ~60.7 & 91 & 19 \\
  ~~8 & 207.573730 & -1.715249 & 36.033 & 35.096 &  110.0 & 0.2293 & 110.6 & -- & 111 \\
  ~~9 & 207.178574 & -1.790581 & 20.194 & 20.500 &  107.4 & 0.1312 & ~44.3 & -- & 52, 175\\
  ~10 & 207.003250 & -1.809448 & 25.858 & 34.192 &  142.0 & 0.1917 & ~69.4 & 1 & 1 \\
  ~11 & 207.019287 & -1.834404 & 31.519 & 44.440 &  104.1 & 0.2413 & 114.2 & 1 & 122 \\
  ~12 & 207.595551 & -1.762271 & 17.131 & 15.834 &  ~79.1 & 0.1062 & ~26.8 & -- & 47 \\
  ~14 & 207.112213 & -1.874567 & 22.331 & 45.132 &  107.9 & 0.2047 & ~67.2 & 18 & -- \\
  ~15 & 207.147217 & -1.880469 & 37.977 & 57.559 &  135.2 & 0.3014 & 135.8 & 66 & 37 \\
  ~19 & 207.080032 & -1.853194 & 15.625 & 41.104 &  127.5 & 0.1634 & ~44.6 & -- & 49, 129 \\
  ~20 & 206.864578 & -1.877663 & 26.845 & 18.099 &  106.5 & 0.1421 & ~31.1 & -- & 21 \\
  ~21 & 207.313156 & -2.535746 & 18.197 & 31.104 &  ~97.6 & 0.1534 & ~32.5 & -- & 69 \\
  ~22 & 207.066986 & -1.820135 & 39.235 & 16.165 &  105.2 & 0.1623 & ~41.0 & -- & 28 \\
  ~23 & 207.292053 & -1.833485 & 28.179 & 30.402 &  ~76.2 & 0.1887 & ~41.1 & -- & 34 \\
  ~24 & 207.325989 & -1.841186 & 35.372 & 41.370 &  ~79.3 & 0.2466 & ~63.1 & -- & 81 \\
  ~25 & 207.539764 & -1.722399 & 20.807 & 23.450 &  138.2 & 0.1424 & ~26.4 & -- & 8 \\
  ~26 & 207.299988 & -2.153447 & 10.363 & 14.769 &  142.4 & 0.0797 & ~11.7 & 4, 74 & 62 \\
  ~28 & 207.387711 & -1.200810 & 22.416 & 55.657 &  126.6 & 0.2277 & ~56.9 & -- & 203 \\
  ~29 & 207.820404 & -1.869027 & 73.438 & 19.956 &  118.3 & 0.2468 & ~67.5 & -- & 87 \\
  ~33 & 206.923126 & -1.625726 & 31.335 & 19.515 &  ~98.0 & 0.1594 & ~28.1 & -- & 140 \\
  ~34 & 207.038132 & -1.823000 & 64.027 & 37.042 &  ~92.7 & 0.3140 & ~78.9 & 1 & 18 \\
  ~38 & 207.403854 & -1.945135 & 28.939 & 61.887 &  110.7 & 0.2728 & ~62.1 & 12 & -- \\
  ~39 & 206.980835 & -1.792985 & 36.982 & 22.868 &  112.6 & 0.1875 & ~31.2 & 7 & 3 \\
  ~40 & 207.299042 & -2.089805 & 30.359 & 37.810 &  ~99.9 & 0.2184 & ~44.3 & -- & 15 \\
  ~41 & 207.068726 & -1.791201 & 33.395 & 47.371 &  106.5 & 0.2564 & ~58.3 & 125 & 104 \\
  ~42 & 207.755325 & -1.936775 & 51.862 & 89.527 &  ~97.4 & 0.4393 & 140.5 & -- & 165 \\
  ~43 & 207.292389 & -2.188894 & 24.373 & 47.070 &  110.0 & 0.2184 & ~47.0 & -- & 30 \\
  ~44 & 207.322525 & -2.136961 & 18.761 & 17.809 &  ~61.7 & 0.1178 & ~14.8 & 4 & -- \\
  ~45 & 207.123840 & -1.894273 & 17.088 & 33.440 &  118.5 & 0.1541 & ~23.1 & 18 & -- \\
  ~46 & 207.213791 & -1.837247 & 20.588 & 20.484 &  162.0 & 0.1324 & ~16.4 & -- & 50 \\
  ~47 & 207.153214 & -1.834679 & 31.781 & 55.356 &  135.8 & 0.2704 & ~55.0 & -- & 48 \\
  ~48 & 207.308899 & -1.805477 & 34.360 & 34.609 &  ~70.6 & 0.2223 & ~40.4 & -- & 99 \\
  ~49 & 207.693375 & -1.925644 & 27.323 & 57.313 &  123.8 & 0.2551 & ~50.2 & -- & 94 \\
  ~51 & 207.172455 & -1.884521 & 51.425 & 42.474 &  ~90.1 & 0.3013 & ~65.4 & -- & 37 \\
  ~52 & 207.020584 & -1.792733 & 41.801 & 40.818 &  ~81.1 & 0.2663 & ~57.7 & 1 & -- \\
  ~56 & 206.945175 & -1.817623 & 20.365 & 21.518 &  ~76.4 & 0.1349 & ~11.2 & 2 & 5 \\
  ~57 & 206.945007 & -1.586929 & 73.244 & 38.464 &  ~96.5 & 0.3422 & ~69.8 & 22 & -- \\
  ~59 & 207.786240 & -1.859325 & 52.540 & 34.661 &  117.2 & 0.2751 & ~51.2 & -- & 106 \\
  ~60 & 206.988373 & -1.821593 & 23.290 & 21.273 &  ~86.1 & 0.1435 & ~14.9 & 33 & 20 \\
  ~61 & 207.227844 & -1.624462 & 73.422 & 35.722 &  117.7 & 0.3302 & ~72.2 & 143 & -- \\
  ~63 & 206.933914 & -1.648822 & 33.063 & 30.730 &  ~78.7 & 0.2055 & ~27.6 & 10 & -- \\
  ~64 & 207.245499 & -1.808280 & 51.262 & 24.315 &  ~97.6 & 0.2276 & ~33.0 & -- & 19 \\
  ~65 & 207.075958 & -1.866452 & 17.369 & 49.979 &  129.9 & 0.1899 & ~23.6 & 102 & 49 \\
  ~66 & 207.431564 & -1.985184 & 95.524 & 76.469 &  151.7 & 0.5511 & 138.4 & 107 & 127 \\
  ~67 & 207.308060 & -1.876241 & 40.374 & 36.242 &  169.4 & 0.2466 & ~35.5 & -- & 44 \\
  ~69 & 207.676498 & -1.556596 & 34.953 & 30.320 &  ~86.9 & 0.2099 & ~27.5 & 262 & 46, 258 \\
  ~71 & 206.790985 & -1.782508 & 35.753 & 58.944 &  106.2 & 0.2960 & ~52.8 & 26 & -- \\
  ~74 & 206.878906 & -1.892263 & 35.983 & 30.359 &  147.9 & 0.2131 & ~27.2 & -- & 262 \\
  ~77 & 207.239197 & -1.577599 & 65.649 & 33.044 &  130.9 & 0.3003 & ~50.4 & -- & 152 \\
  ~78 & 206.924118 & -1.812391 & 48.701 & 37.169 &  ~~3.5 & 0.2743 & ~40.4 & 2 & -- \\
  ~80 & 206.957672 & -1.794173 & 45.861 & 27.254 &  126.4 & 0.2279 & ~29.8 & 2 & -- \\
  ~84 & 207.366516 & -1.951482 & 44.769 & 85.544 &  126.0 & 0.3990 & ~87.4 & -- & 72 \\
  ~89 & 207.027054 & -1.849236 & 14.666 & 20.791 &  134.8 & 0.1126 & ~~8.2 & 60 & 122 \\
  ~92 & 207.430634 & -1.951102 & 18.764 & 16.838 &  153.5 & 0.1146 & ~~7.6 & 34 & 9 \\
  ~95 & 207.835007 & -1.837569 & 26.368 & 49.742 &  127.3 & 0.2335 & ~29.8 & -- & 87 \\
  ~97 & 207.774399 & -1.914737 & 57.055 & 41.598 &  ~89.2 & 0.3141 & ~43.9 & -- & 165 \\
  ~98 & 207.109055 & -1.810394 & 34.920 & 27.630 &  173.2 & 0.2002 & ~21.5 & -- & 65 \\
  101 & 207.304947 & -1.823945 & 18.227 & 28.587 &  ~90.6 & 0.1471 & ~12.5 & -- & 34 \\
  104 & 207.786575 & -1.771225 & 66.501 & 44.612 &  ~92.0 & 0.3512 & ~52.0 & -- & 54 \\
\hline 
\hline 
\end{tabular} 
\end{center}
\end{table*}

\begin{table*}
\contcaption{~} 
\label{table_Dust_assoc_clumps_continued} 
\begin{center}
\begin{tabular}{r@{~~}c@{~~}c@{~~}c@{~~}c@{~~}c@{~~}crcc}
\hline 
\hline 
\# & $l$ & $b$ & Axis 1 & Axis 2 & Orient. angle & Size & Mass & \twelveCO~ass. & \thirthCO~ass. \\
~  & [deg] & [deg] & [\arcsec] & [\arcsec] & [deg] & [pc] & [\Msol] & ~ & ~ \\
\hline
107 & 206.906769 & -1.794024 & 35.791 & 19.090 & 121.8 & 0.1685 & ~14.6 & 23 & 25 \\
108 & 207.329880 & -1.902573 & 40.578 & 62.026 & ~95.5 & 0.3234 & ~45.3 & -- & 148 \\
109 & 207.317841 & -1.849597 & 13.726 & 39.271 & 129.8 & 0.1497 & ~12.0 & -- & 81 \\
111 & 207.274780 & -2.146296 & 54.046 & 48.001 & 148.7 & 0.3284 & ~52.8 & -- & 7 \\
117 & 207.333359 & -2.171435 & 33.985 & 27.641 & 113.9 & 0.1976 & ~20.0 & 4 & 2 \\
120 & 207.339050 & -2.569749 & 50.174 & 44.283 & 103.8 & 0.3039 & ~40.6 & -- & 202 \\
124 & 207.688248 & -1.648921 & 17.702 & 23.262 & 140.1 & 0.1308 & ~~9.2 & -- & 189 \\
128 & 207.292252 & -2.540013 & 42.094 & 41.456 & 162.6 & 0.2693 & ~29.5 & -- & 69 \\
130 & 207.655762 & -1.907992 & 27.191 & 38.539 & ~99.2 & 0.2087 & ~18.7 & -- & 158 \\
131 & 206.931473 & -1.609767 & 21.948 & 25.764 & 155.5 & 0.1533 & ~~9.2 & 22 & 85 \\
133 & 207.401382 & -1.383992 & 62.297 & 36.012 & 117.1 & 0.3054 & ~37.9 & -- & 143 \\
134 & 207.165543 & -1.806305 & 29.583 & 44.250 & ~98.6 & 0.2332 & ~22.5 & -- & 175 \\
137 & 207.550400 & -1.708699 & 17.785 & 20.706 & 168.7 & 0.1237 & ~~6.8 & -- & 8 \\
139 & 207.588120 & -1.730466 & 29.901 & 20.575 & 138.7 & 0.1599 & ~11.0 & -- & 93 \\
148 & 206.881012 & -1.801407 & 45.481 & 22.142 & 120.8 & 0.2046 & ~17.0 & 23 & 109 \\
149 & 207.083038 & -1.882143 & 16.431 & 34.357 & 117.0 & 0.1532 & ~12.1 & 102 & 49 \\
153 & 206.857147 & -1.738360 & 98.238 & 56.385 & 138.7 & 0.4799 & ~70.8 & 158 & -- \\
160 & 207.299332 & -1.841808 & 10.883 & 12.767 & ~87.6 & 0.0760 & ~~4.0 & -- & 34 \\
163 & 207.114883 & -1.860419 & 14.523 & 21.385 & ~~1.7 & 0.1136 & ~~7.9 & 18 & -- \\
166 & 207.314865 & -2.181952 & 21.313 & 30.151 & 137.9 & 0.1634 & ~12.0 & 4 & 116 \\
168 & 207.549973 & -1.738834 & 36.370 & 27.033 & 137.5 & 0.2021 & ~13.9 & -- & 100 \\
169 & 206.940979 & -1.558514 & 12.349 & 15.210 & 127.0 & 0.0883 & ~~4.3 & 221 & 45 \\
171 & 207.356705 & -1.850779 & 15.648 & 40.200 & 116.2 & 0.1617 & ~10.8 & -- & 91 \\
177 & 206.973099 & -1.815511 & 18.204 & 20.166 & ~84.9 & 0.1235 & ~~5.7 & 2 & -- \\
180 & 207.363998 & -1.916371 & 32.295 & 31.268 & ~69.6 & 0.2049 & ~13.8 & -- & 96 \\
183 & 207.192856 & -1.782771 & 31.377 & 17.568 & ~92.0 & 0.1513 & ~~7.5 & -- & 52 \\
187 & 206.929871 & -1.855065 & 30.898 & 28.633 & ~85.4 & 0.1917 & ~12.5 & 8 & -- \\
194 & 207.164444 & -1.856172 & 17.636 &  8.477 & 110.9 & 0.0788 & ~~3.7 & -- & 59 \\
195 & 207.836929 & -1.851832 & 16.045 & 23.894 & ~83.5 & 0.1262 & ~~4.4 & -- & 87 \\
198 & 207.270752 & -2.433175 & 38.737 & 29.128 & 151.7 & 0.2165 & ~~9.9 & -- & 51 \\
201 & 207.327957 & -2.293849 & 26.639 & 34.908 & ~99.4 & 0.1966 & ~11.9 & 200 & 98, 275 \\
205 & 207.790039 & -1.894972 & 47.352 & 41.698 & 167.4 & 0.2865 & ~20.7 & -- & 106 \\
207 & 207.376190 & -2.363848 & 16.868 & 31.244 & 114.7 & 0.1480 & ~~7.2 & -- & 248 \\
208 & 207.273651 & -1.873285 & 16.118 & 31.808 & 112.0 & 0.1460 & ~~7.9 & -- & 79 \\
211 & 207.132492 & -1.841401 & 15.694 & 12.398 & 136.1 & 0.0899 & ~~5.3 & -- & 6 \\
219 & 206.902893 & -1.671969 & 41.576 & 69.605 & 122.9 & 0.3468 & ~34.7 & 218 & -- \\
222 & 206.899124 & -1.780821 & 25.062 & 17.500 & 137.2 & 0.1350 & ~~6.3 & -- & 25 \\
224 & 206.945557 & -1.662166 & 28.658 & 18.967 & 135.2 & 0.1503 & ~~7.0 & 10 &  -- \\
236 & 206.842758 & -1.805897 & 19.914 & 21.283 & 110.3 & 0.1327 & ~~5.6 & 119 & --  \\
239 & 206.943497 & -1.796287 & 11.929 & 15.309 & ~88.9 & 0.0871 & ~~2.7 & 2 & -- \\
241 & 207.338394 & -2.346179 & 37.713 & 32.027 & 151.1 & 0.2240 & ~13.2 & 200 & 103 \\
242 & 207.259857 & -1.794912 & 39.910 &  5.616 & 118.8 & 0.0965 & ~~5.8 & -- & 19 \\
243 & 207.343353 & -1.952700 & 18.851 & 20.779 & 157.1 & 0.1276 & ~~4.9 & -- & 71 \\
249 & 206.788879 & -1.764879 & 31.999 & 44.055 & ~82.5 & 0.2421 & ~10.7 & 26, 227 & --\\
251 & 207.115112 & -1.642861 & 22.701 & 24.356 & 166.6 & 0.1516 & ~~6.9 & 152 & 57 \\
257 & 206.921402 & -1.878695 & 42.927 & 20.147 & 104.9 & 0.1896 & ~~9.9 & 37 & 124 \\
259 & 207.476700 & -1.695197 & 29.006 & 24.083 & 131.9 & 0.1704 & ~~8.6 & -- & 216 \\
260 & 207.238800 & -1.748492 & 23.584 & 22.739 & 155.5 & 0.1493 & ~~6.9 & -- & 126 \\
261 & 207.288773 & -1.800261 & 13.921 & 27.102 & ~90.8 & 0.1252 & ~~4.9 & -- & 99 \\
265 & 207.293915 & -1.754439 & 67.400 & 30.956 & 109.9 & 0.2945 & ~25.7 & -- & 223 \\
271 & 207.067841 & -1.845766 & 13.249 &  9.539 & 139.7 & 0.0724 & ~~3.8 & 103 & -- \\
273 & 207.915878 & -1.804306 & 21.767 & 32.740 & ~51.9 & 0.1721 & ~~7.6 & -- & 150 \\
281 & 207.106552 & -1.873314 & 12.193 & 14.163 & 170.8 & 0.0847 & ~~3.8 & 42 & -- \\
282 & 207.020798 & -1.817548 & 15.157 & 11.241 & ~83.5 & 0.0841 & ~~2.9 & 1 & 1 \\
286 & 206.995224 & -1.783564 & 26.814 & 21.207 & ~82.3 & 0.1537 & ~~5.7 & 7, 81 & 3 \\
292 & 207.155762 & -1.817985 & 14.252 & 14.853 & ~73.4 & 0.0938 & ~~2.9 & -- & 48 \\
293 & 207.457932 & -1.966992 & 18.410 & 34.791 & ~46.2 & 0.1631 & ~~6.3 & 64 & 251 \\
294 & 207.313507 & -1.796444 &  9.083 &  8.367 & ~66.6 & 0.0562 & ~~1.8 & -- & 99 \\
\hline 
\hline 
\end{tabular} 
\end{center}
\end{table*}

\begin{table*}
\contcaption{~} 
\label{table_Dust_assoc_clumps_continued_ii} 
\begin{center}
\begin{tabular}{r@{~~}c@{~~}c@{~~}c@{~~}c@{~~}c@{~~}crcc}
\hline 
\hline 
\# & $l$ & $b$ & Axis 1 & Axis 2 & Orient. angle & Size & Mass & \twelveCO~ass. & \thirthCO~ass. \\
~  & [deg] & [deg] & [\arcsec] & [\arcsec] & [deg] & [pc] & [\Msol] & ~ & ~ \\
\hline
305 & 207.000534 & -1.932781 & 22.117 & 22.148 & ~71.2 & 0.1427 & ~~5.9 & 194 & -- \\
306 & 207.373520 & -1.771336 & 24.079 & 24.061 & 161.5 & 0.1552 & ~~5.9 & -- & 134 \\
318 & 207.821960 & -1.855166 &  7.031 & 26.967 & 106.3 & 0.0887 & ~~3.0 & -- & 87 \\
319 & 207.139359 & -1.871427 & 44.644 & 13.757 & 106.2 & 0.1598 & ~~8.1 & -- & 11 \\
328 & 206.912903 & -1.851761 & 40.046 & 11.939 & 110.5 & 0.1409 & ~~5.4 & 57 & -- \\
336 & 207.214905 & -1.859819 & 24.036 & 25.808 & ~77.7 & 0.1606 & ~~6.1 & -- & 92 \\
337 & 207.297348 & -2.131625 & 13.907 & 14.396 & 170.2 & 0.0912 & ~~3.4 & 308 & 35 \\
338 & 207.126709 & -1.613373 & 19.227 & 21.244 & 113.5 & 0.1303 & ~~4.6 & 376 & -- \\
346 & 207.263123 & -1.774051 & 13.282 & 40.227 & 119.8 & 0.1490 & ~~6.6 & -- & 114 \\
358 & 207.184021 & -1.774454 &100.202 & 88.068 & ~67.8 & 0.6057 & ~19.3 & -- & 52 \\
359 & 207.143311 & -1.805941 & 25.185 & 22.923 & 176.9 & 0.1549 & ~~6.3 & -- & 48 \\
360 & 207.071030 & -1.800395 & 12.081 & 26.400 & ~~1.9 & 0.1151 & ~~2.7 & -- & 28, 104 \\
364 & 207.289032 & -1.832601 &  8.868 & 13.079 & ~18.5 & 0.0694 & ~~2.1 & -- & 34 \\
366 & 207.270325 & -1.839894 & 15.744 & 26.471 & 105.5 & 0.1316 & ~~5.2 & -- & 79 \\
\hline 
\hline 
\end{tabular} 
\end{center}
\end{table*}

\subsection{Test of the optical depth correction}
\label{appx_sigma_vel}

\begin{figure} 
\includegraphics[width=84mm]{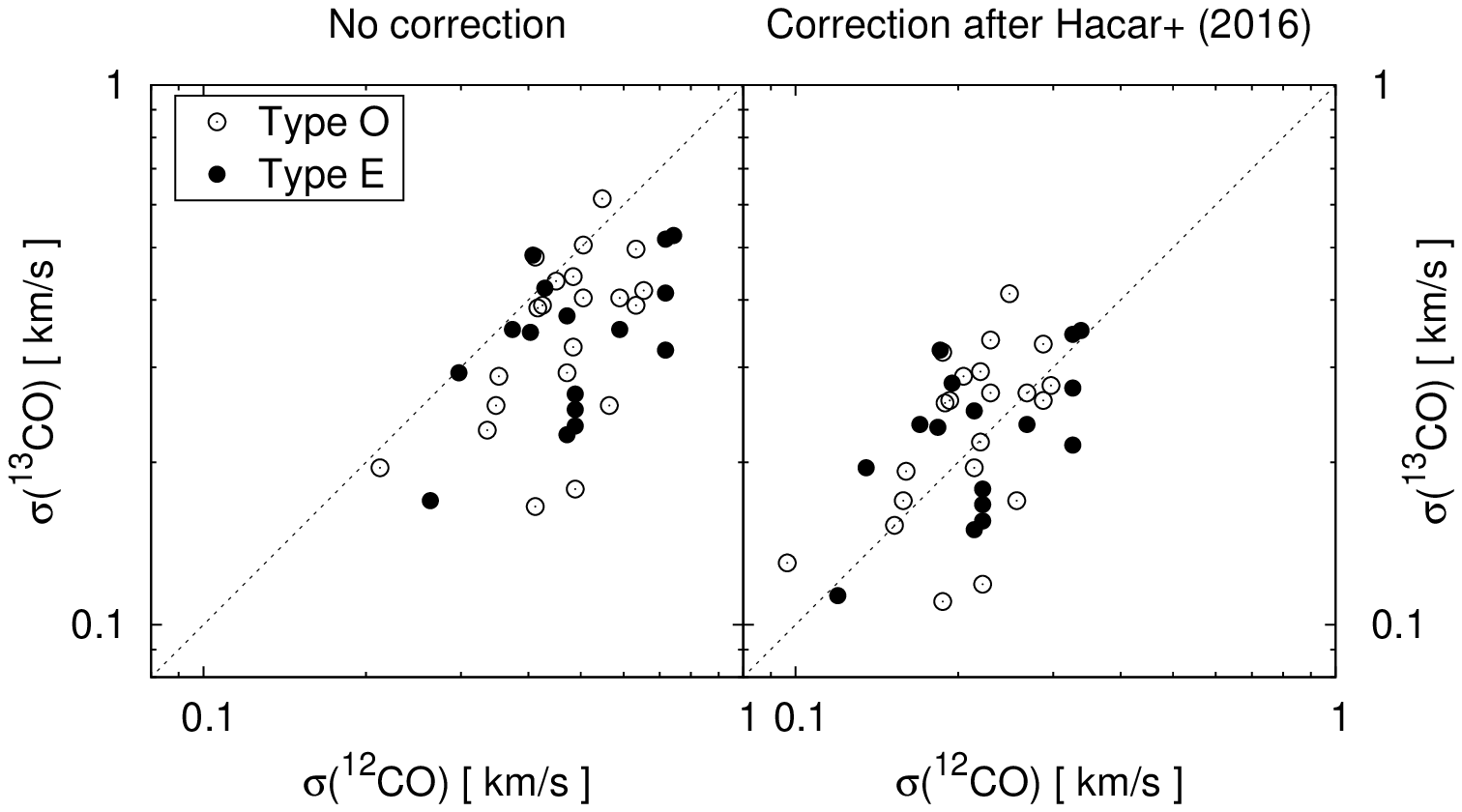}
\vspace{0.5cm}  
\caption{Comparison of the velocity dispersion measured in \twelveCO{} and \thirthCO{} for associated clumps. The left plot compares the directly measured line widths, the right plot shows the dispersions after the optical depth correction.}
\label{fig_sigma-sigma}
\end{figure}

To measure the velocity dispersion of molecules from the line width of optically thick lines we have to apply an optical depth broadening following \citet{Hacar2016}. Using the measured $X$-factors from in Sect.~\ref{Estimation of X-factor} we derive in Sect.~\ref{sect_size-linewidth} approximate correction factors of about 2.2 for the \twelveCO{} and about 1.4 for the \thirthCO{} clumps. Both factors are uncertain within about 10\,\% as the computation of the optical depth from the $X$-factors tends to underestimate the optical depth, due to the neglect of the change of the line shape, while the use of the relation for the thermal line width from \citet{Hacar2016}, ignoring the small suprathermal contribution (see Sect.~\ref{Virial analysis}), tends to overestimate the correction.  To verify the correction factors for the two linewidths we show in Fig.~\ref{fig_sigma-sigma} the velocity dispersion measured in \twelveCO{} and \thirthCO{} for associated clumps that should see the same gas. Without the optical depth correction there is a clear offset from the identity line. After the correction both velocity dispersions match within a factor of two, symmetric to the identity line. This indicates that we obtain reliable estimates of the clump velocity dispersions.

\end{document}